\documentclass[11pt]{article}
\usepackage[latin1]{inputenc}
\usepackage{amsmath}
\usepackage{amsfonts}
\usepackage{amssymb}
\usepackage{color}
\usepackage{verbatim}

\usepackage{epsfig}
\newlength{\bredde}
\def\slash#1{\settowidth{\bredde}{$#1$}\ifmmode\,\raisebox{.15ex}{/}
\hspace*{-\bredde} #1\else$\,\raisebox{.15ex}{/}\hspace*{-\bredde} #1$\fi}
\newcommand{\be}{\begin{equation}}
\newcommand{\ee}{\end{equation}}
\newcommand{\bea}{\begin{eqnarray}}
\newcommand{\eea}{\end{eqnarray}}

\newcommand{\la}{\lambda}
\newcommand{\al}{\alpha}

\newcommand{\bs}{\begin{split}}
\newcommand{\es}{\end{split}}

\newcommand{\sect}[1]{\setcounter{equation}{0}\section{#1}}

\def\Tr{{\mbox{Tr}}}

\begin{document}
\topmargin -1.4cm
\oddsidemargin -0.8cm
\evensidemargin -0.8cm
\title{\Large\bf
Universal Correlations and Power-Law Tails\\
in Financial Covariance Matrices
}

\vspace{1.5cm}
\author{~\\{\sc G.~Akemann}$^1$, {\sc J.~Fischmann}$^2$ and
{\sc P.~Vivo}$^3$
\\~\\
$^1$Department of Mathematical Sciences \& BURSt Research Centre\\
Brunel University West London,
Uxbridge UB8 3PH,
United Kingdom\\~\\
$^2$School of Mathematical Sciences\\
Queen Mary University of London, London E1
4NS, United Kingdom
\\~\\
$^3$Abdus Salam International Centre for Theoretical Physics\\
Strada Costiera 11, 34014 Trieste, Italy
}

\date{}
\maketitle
\vfill
\begin{abstract}
Signatures of universality are detected by comparing individual eigenvalue
distributions and level spacings from financial
covariance matrices to random matrix predictions.
A chopping procedure is devised in order to produce a statistical ensemble of
asset-price covariances from a single instance of financial data sets. Local
results for the smallest eigenvalue and individual spacings are very stable
upon reshuffling the time windows and assets. They are in good agreement
with the universal Tracy-Widom distribution and Wigner surmise, respectively.
This suggests a strong degree of robustness especially in the low-lying sector
of the spectra, most relevant for portfolio selections.
Conversely, the global spectral density of a single covariance matrix
as well as the average over all unfolded nearest-neighbour
spacing distributions deviate from standard Gaussian random matrix
predictions.  The data are in fair agreement
with a recently introduced generalised random matrix model,
with correlations showing a power-law decay.

\end{abstract}

\vfill

\thispagestyle{empty}
\newpage

\renewcommand{\thefootnote}{\arabic{footnote}}
\setcounter{footnote}{0}

\sect{Introduction}\label{intro}

'Econophysics', a hybrid of 'economy' and 'physics', has become a very active
field of research in the past decade. In this paper we will focus on aspects
of portfolio selection and market modelling. In particular we will analyse
cross-correlations in financial time-series and compare to the predictions
of Random Matrix Theory (RMT). We refer to the reviews
\cite{PBLreview,burdareview} on this subject, as well as for
other methods e.g. from field theory to \cite{sornette}.

A still debated issue in the community is to what extent the 'historical'
determination of covariance estimators (i.e. based on past time-series over a
\emph{finite} temporal window $T$) can be trusted when forecasting the
financial risk of a certain portfolio; put it differently, how reliably is the
past going to shape the future? In a pioneering paper, Laloux {\it et al.}
\cite{laloux} used a comparison with RMT to cast serious doubts on the
usefulness of historical covariance spectra
in estimating the variance (and thus the future risk) of a given portfolio,
questioning the widely applied
procedure of Markowitz's theory  based on Gaussian mean-field approximations.
The 'measurement noise' due to the finiteness of the historical time series $T$
was claimed in \cite{laloux} to bury most of the relevant information encoded
in the historical covariance matrices,
thus impairing \emph{ab ovo} much of the consequent predictions.
In subsequent papers the work of \cite{laloux} was repeated and refined to
other quantities, using Gaussian RMT
\cite{plerouplus,Plerouetal,lillo,utsugi}, non-Gaussian heavy tailed
distributions \cite{galluccio,burdang,Bphysics,biroli,AV,AAV,germano},
or RMT with complex
eigenvalues using lagged time series \cite{Kwapien,BT}. In addition
clever methods were devised to detect meaningful correlations buried under
the 'noise-dressed' regions of the spectra \cite{GK,burdanoise}, thus
trying to mitigate the pessimistic forecast of \cite{laloux}.

A central tool for our
comparison with empirical financial data is the so-called
Wishart-Laguerre (WL) ensemble of random matrices.
The WL ensemble contains random matrices of the form \cite{wishart}
$\mathbf{W}=(1/T)\mathbf{X}^{\dagger}\mathbf{X}$, where
$\mathbf{X}$ is a rectangular matrix of size $T\times N~$($T>N$),
whose entries are independent Gaussian variables in the simplest
case, and $\mathbf{X^{\dagger}}$ is the Hermitian conjugate of
$\mathbf{X}$. As such, the WL ensemble contains (positive
definite) covariance matrices $\mathbf{W}$ of maximally random
data sets. They have since appeared in many different contexts apart from
quantitative finance,
ranging from mathematical statistics \cite{IMJ} and statistical physics
\cite{stat}
to gauge theories \cite{gauge}, quantum gravity \cite{QG} and
telecommunications \cite{tele}.

By definition, the WL ensemble constitutes a 'null hypothesis',
with the highest degree of randomness
and lowest degree of built-in information in the data-set, and therefore is
an ideal benchmark for comparison
with the \emph{a priori} much richer correlations in the financial time
series.

Within the large-$N$ predictions of WL one has to distinguish
between \emph{global} spectral properties, taking into account
correlations over a scale of $O(1)$ much larger than the mean level spacing,
and \emph{local} properties, probing
correlations on the scale of the mean spacing (typically of $O(1/N)$).
It is well known that
the latter are much more robust under small deformations of the
Gaussian probability distribution of the matrix elements. Typical examples
include the distribution of individual eigenvalues and individual spacings
among eigenvalues in the bulk.

While the global spectral density has received a good deal of attention
in the works quoted above, the spacing distribution has so far only been
investigated in \cite{plerouplus,Plerouetal}, histogramming the spacings
between all consecutive eigenvalues after unfolding.
One of the aims of this paper is to further investigate the \emph{local}
statistics (individual distributions),
taking also into account the presence of non-Gaussian heavy tails in the
spectral density.

In order to test e.g. if the smallest eigenvalue follows the universal
Tracy-Widom distribution \cite{TW}, as was proven only very recently for WL
with real matrix elements \cite{Sodin}, we immediately face the
following problem: a single covariance matrix clearly does not allow for
testing individual eigenvalue distributions.
We are therefore led to define a meaningful way of generating
ensembles of covariance matrices from a single data set. These are then tested
against WL and its generalisation with power-law tails that was
proposed in
\cite{BGW} and further developed for local statistics in \cite{AV} as well as
in this paper.

This article is organised as follows. In section \ref{rmt}
we summarise
the relevant predictions from RMT, both for the standard WL ensemble and its
non-Gaussian generalisation with power-law tails.
In section \ref{data} we compare these to data from financial covariance
matrices. This section is divided into two parts. In \ref{global}
we compare to the data
from a single covariance matrix partly repeating
previous analysis. In the second subsection \ref{indiv}
we define ensembles
of matrices by chopping the covariance matrix, and compare to the distribution
of individual eigenvalues and spacings.
In section \ref{con} we offer some conclusions, followed by 3 appendices
containing technical details and consistency checks.

\sect{Random Matrix Predictions}\label{rmt}

In this section we summarise the analytical predictions from RMT, both for the
standard Gaussian model and its generalisation. We do
not give derivations here but rather illustrate these predictions by
comparing to numerically generated random matrices for both models.

\subsection{Global density and level spacing}

We first introduce the standard Wishart-Laguerre (WL) ensemble of Gaussian
random
matrices (also called chiral Gaussian ensemble).
In view of our application to time series, in particular
stock price fluctuations, we restrict ourselves to
real matrix elements denoted by the Dyson index $\beta=1$.
Its probability density distribution of matrix elements
is given by
\begin{equation}
\label{PWL}
{\cal P}_{{W\!L}}\left( \mathbf{X}\right)
\sim\exp\left[ -\sigma^2\Tr(\mathbf{X}^{T}\mathbf{X})  \right] ,
\end{equation}
where \mbox{\bf X} is a matrix of size $T\times N~$ ($T>N$)
with real elements.
The integration measure $d\mathbf{X}$ is defined by
integrating over all independent matrix elements of \mbox{\bf X}
with a flat measure. The joint probability distribution
of the positive definite eigenvalues of
$\mathbf{W}=(1/T)
\mathbf{X^{\dagger}}\mathbf{X}$ (the so-called Wishart matrix), or
equivalently the singular values of the matrix $\mathbf{X}$ is obtained by
integrating over the angular degrees of freedom.
Since we are only interested in two specific
correlation functions in the large-$N$ limit
we give the results without derivation.

The global spectral density defined as $\rho(\la)=\langle
\Tr~\delta(\la-\mathbf{W})\rangle$,  averaged
with respect to (\ref{PWL}). In the large-$N$ limit it is given by  \cite{MP}
\be
\label{MP}
\rho_{M\!P}(x) =\ \frac{1}{2\pi c x}
\sqrt{(x-cX_-)(cX_+-x)}\ ,\ \ \mbox{with}\ \ x\in[cX_-,cX_+]\ .
\ee
This is called Mar\v{c}enko-Pastur law, and we have
normalised it to have
integral and first moment equal to unity\footnote{In previous comparisons to
  data the 
inverse  variance $\sigma$ of the distribution eq. (\ref{PWL}) was used as a
free parameter.}  (see e.g. fig \ref{Comparison}).
The endpoints of support are
\be
X_\pm\ \equiv\ (c^{-\frac12}\pm1)^2\ ,
\ee
where both the large-$N$ and large-$T$ limit is taken with $N=cT$, keeping
$0<c\leq 1$. Indeed, the applications to
times series analysis require $T$ to be much larger than the number of stocks
$N \ll T$.

A second quantity of interest is the spacing distribution between consecutive
levels in the bulk of the spectrum,
probing correlations on a local level of order $O(1/N)$.
We make a distinction between \emph{global} and
\emph{individual} nearest-neighbour distributions.
The former is obtained by averaging over spacings in the bulk after unfolding
(see appendix \ref{unfold}) and is commonly used in comparisons to RMT.
The latter is defined as the distribution of the spacing $s_k$ between the
$k$th and $(k-1)$st eigenvalue, for fixed $k$ and $N$
(see e.g. \cite{muller}):
\begin{equation}\label{individual}
 s_k=\frac{\lambda_k-\lambda_{k-1}}{\langle
\lambda_k-\lambda_{k-1}\rangle}\ ,
\end{equation}
and it provides
information about local correlations at a given point in the bulk, without
making any further assumptions.

Within RMT the global and individual spacing in the bulk agree, and
an excellent approximation for both distributions can be obtained by applying
the so-called Wigner surmise (WS), based on a $2\times2$ matrix calculation. As
was explained and
checked numerically in \cite{AAV},
the correct surmise for WL is obtained from
the Wigner-Dyson (or Gaussian) ensembles (see also
\cite{plerouplus,Plerouetal})
\be
p_{W\!D}(s)= \frac\pi2 s\ \exp\left(-\frac\pi4 s^2
\right)\ ,
\label{WD}
\ee
here for $\beta=1$. It has norm and first moment of unity. Eqs. (\ref{MP}) and
(\ref{WD}) are very well known and have been compared to financial data
in previous works (see e.g. \cite{laloux,lillo,utsugi,DKO}
and \cite{plerouplus,Plerouetal} respectively).

We now turn to the recent generalisation of WL
introduced in \cite{BGW,AV}.
It can be derived from
a superstatistical or generalised entropy approach,
representing an interpolation between the fully chaotic or random statistics of
the Gaussian WL, and a regular or integrable behaviour.
The generalisation of the
probability distribution of matrix elements eq. (\ref{PWL})
reads
\be
\label{Pgen}
{\cal P}_\al\left( \mathbf{X}\right)
\sim \left( 1+\frac{1}{\al+\frac12 NT+1}
\Tr(\mathbf{X}^{T}\mathbf{X})\right)^{-(\al+\frac12 NT+1)} ,\ \ \al>0\ ,
\ee
where the $N$-dependence is required by convergence.
The resulting global density that generalises the
Mar\v{c}enko-Pastur
density eq. (\ref{MP}) is reading \cite{BGW}
\be
\rho_\al(x)\ =\ \frac{1}{2 \pi c \al \Gamma(\al+1)}
\left(\frac{c\al}{x}\right)^{\al+2} \int_{X_-}^{X_+}dt\,t^{\al}
\exp\left[-\frac{c\al}{x}\,t\right] \sqrt{(t-X_-)(X_+-t)}\ ,
\label{rhogen}
\ee
and it displays a power law tail \cite{AV}.
The density is normalised to unity and has first
moment equal to one (see fig. \ref{Comparison}
for illustration).  The integral can be
computed in terms of a confluent hypergeometric
series \cite{AV}.
For large arguments the density decays algebraically as
$\rho_\al(x\gg1)\sim x^{-\al-2}$. For small arguments the density
is suppressed exponentially, and
for a more detailed discussion
we refer to \cite{AV}. The standard WL quantities are recovered when
$\alpha\to\infty$.

In complete analogy with the standard WL ensemble,
we can define the spacing distribution for the generalised ensemble
eq. (\ref{Pgen}) that generalises eq. (\ref{WD}). It is
based on a Wigner surmise for the corresponding generalised WD ensemble
(see also \cite{Toscanoetc}).
Although this quantity was not derived in \cite{AV}, it
follows in analogy to the corresponding quantity in \cite{AAV}
where generalisations of \eqref{WD} with (stretched)
exponential tails were considered. We shall
therefore be brief and only quote the answer for $\beta=1$ relevant here:
\be
\label{genspace}
p_\al(s) =
\frac{\pi s}{2\al^2\Gamma(\al+1)}\int_0^\infty dt\, t^{\al+2}
\exp\left[ -t-\frac{\pi t^2}{4\al^2}\,s^2\right]\ .
\ee
The norm and first moment are again chosen to be unity. Exploiting the solution
of the integral in terms of Hypergeometric functions it is easy to see that
the local spacing distribution decays with the same power as the global
density eq. (\ref{rhogen}),
$p_\al(s\gg1)\sim s^{-\al-2}$.  A saddle point evaluation
confirms that \eqref{genspace} reproduces \eqref{WD} as expected when
$\al\to\infty$.

\begin{center}
\begin{figure*}[h]
  \unitlength1.0cm
\centerline{\epsfig{file=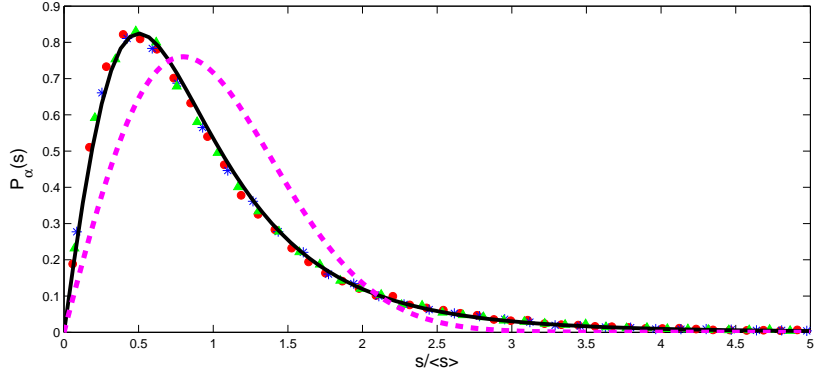,clip=,width=12.8cm}}
  \caption{
\label{spacingtest}
Test of the large-$N$ individual
spacing distribution in the generalised WL model
eq. \eqref{genspace} (full line) vs. the individual
spacing distributions \eqref{individual} for $k=4,6,8$ at $N=8$, $T=16$ and
$\alpha=3$,  from
numerically generated random matrices with distribution \eqref{Pgen}.
The spacing distribution for the standard WL Eq. (\ref{WD})
is added for comparison (pink dashed line).
}
\end{figure*}
\end{center}

We have verified numerically
that the generalised spacing distribution (\ref{genspace})
indeed agrees with individual spacings (eq. \eqref{individual}) from random
matrices from eq. 
\eqref{Pgen}. This is shown in fig. \ref{spacingtest}.

In both ensembles the results given so far, the global density and the global
spacing, can be tested on a single realization of a random or data
matrix.
For individual spacings as well as for the distribution of individual
eigenvalues we need an ensemble of matrices, and the
predictions for the latter are given in the next subsection.


\subsection{Smallest and largest eigenvalues}\label{small}

For the Gaussian WL ensemble with real matrix entries, both the distributions
of the largest and smallest eigenvalue have been rigorously studied by
Johnstone \cite{IMJ}, and very recently
by Feldheim and Sodin \cite{Sodin}, respectively.
In a well defined scaling limit
both cases follow a Tracy-Widom (TW) \cite{TW} distribution
$F_1(s)$.
More precisely, there exist $N$-dependent constants
$a_N^{(L)},b_N^{(L)},a_N^{(S)},b_N^{(S)}$ such that
\be
\mbox{Prob}[\chi_{{\rm min} ( {\rm max})}<x]=F_1(x)
\label{prob}
\ee
where we have introduced the following scaled random variables:
\begin{align}
\chi_{{\rm min}} &=\ \frac{a_N^{(S)}-\la_{{\rm min}}}{b_N^{(S)}}\ ,
\label{min}\\
\chi_{{\rm max}} &=\ \frac{\la_{{\rm max}}-a_N^{(L)}}{b_N^{(L)}}\ .
\label{max}
\end{align}
The Tracy-Widom distribution $F_1(x)$ defined in eq. (\ref{prob}) is given by
\be
F_1(x)\equiv \exp\left[ -\int_x^\infty ds \Big(q(s)+(s-x)q(s)^2\Big)\right]
\label{TW1}
\ee
where $q(s)$ is the solution to the second Painlev\'e equation (PII)
\be
q(s)''=sq(s)+2q(s)^3
\label{PII}
\ee
subject to the boundary condition of approaching the Airy function
asymptotically,
$q(s)\sim \mbox{Ai}(s)$ for $s\to\infty$.
This is also called the Hastings-McLeod solution of PII.

\begin{center}
\begin{figure*}[h]
  \unitlength1.0cm
\centerline{
\epsfig{file=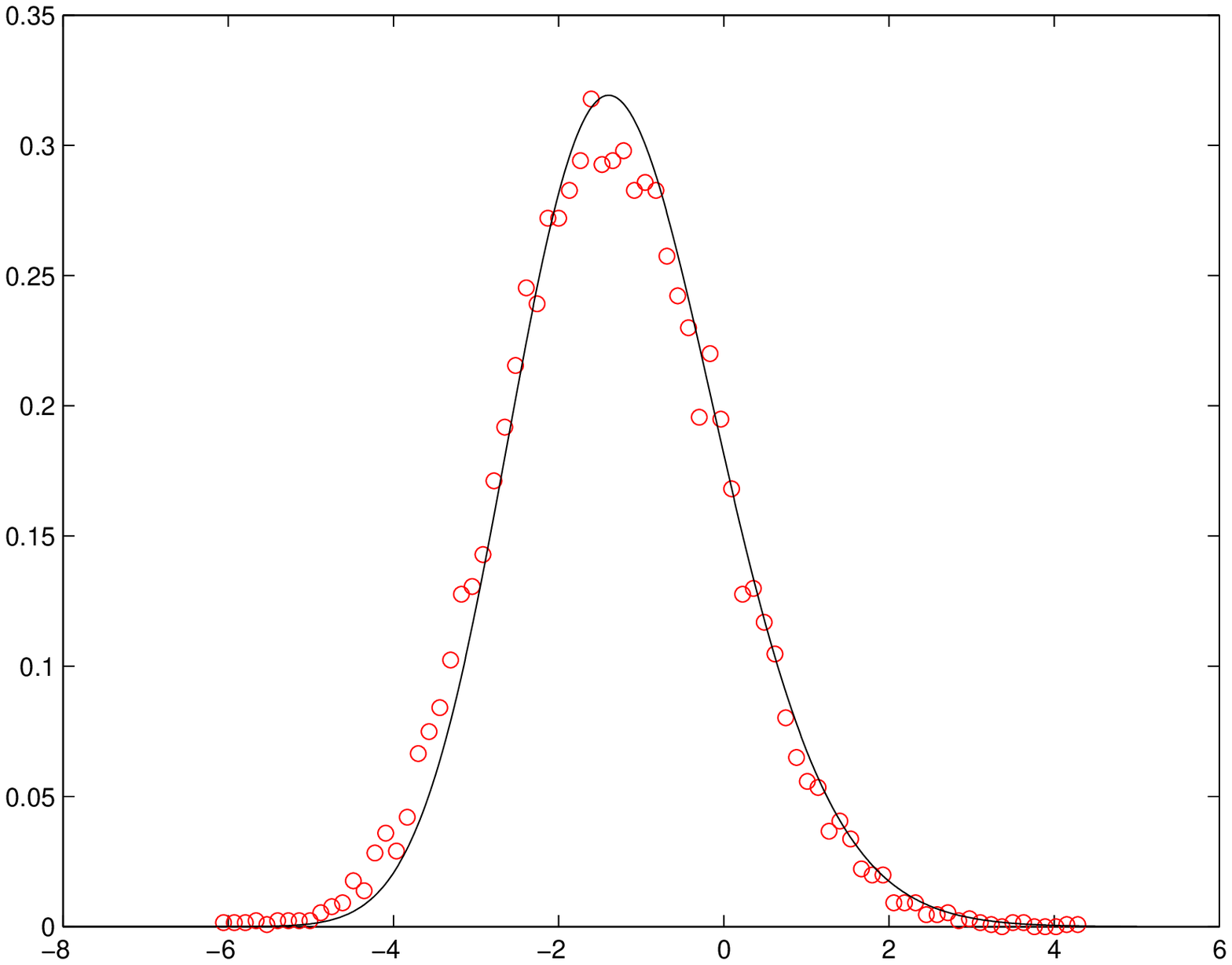,clip=,width=5.8cm}
\epsfig{file=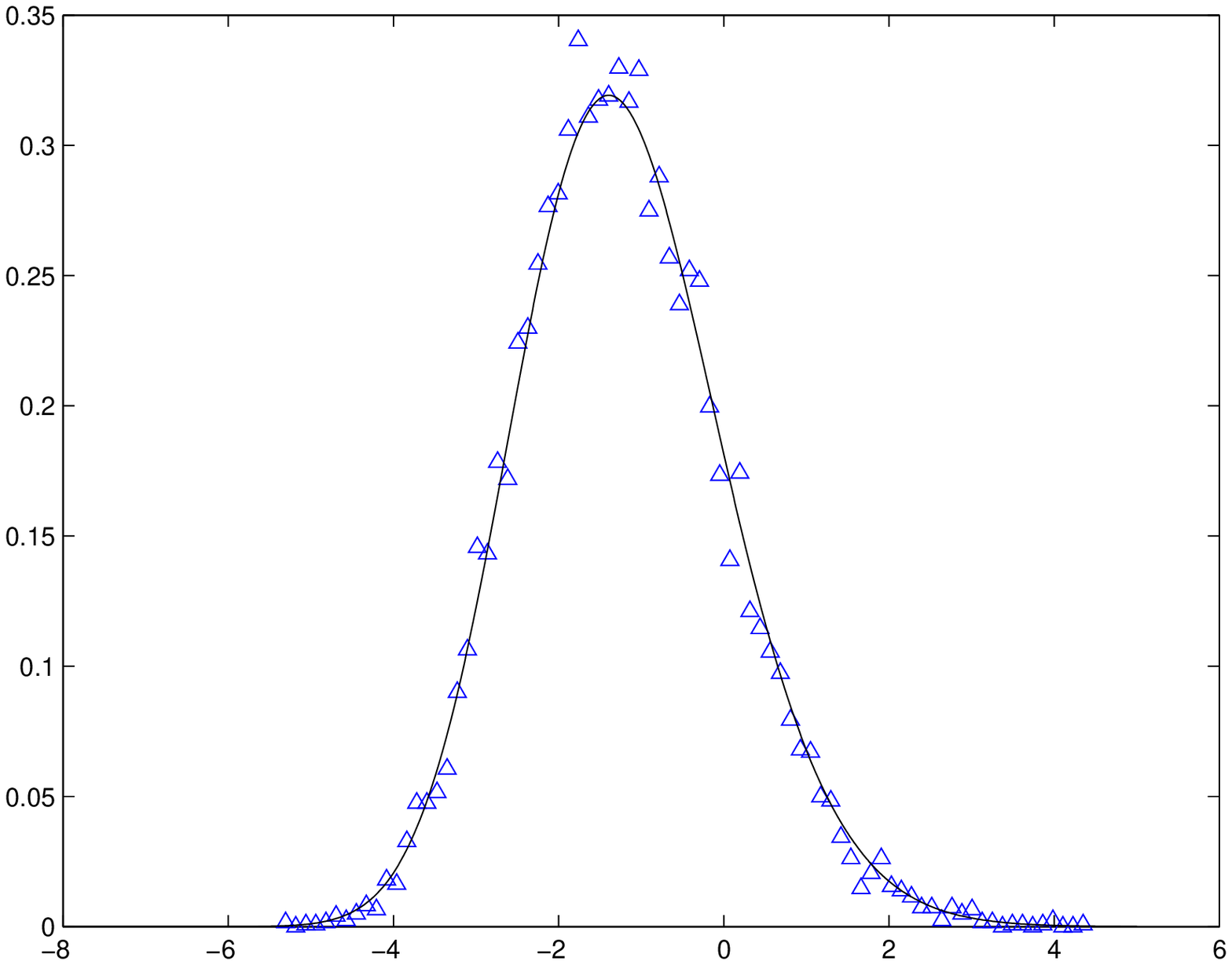,clip=,width=5.8cm}
}
  \caption{
    \label{TWnumcheck}
The distribution of the smallest (left) and largest (right) eigenvalues for
$\beta=1$ WL, after centring and the rescalings eqs. (\ref{min}) and
(\ref{max}). 
The full line corresponds to the
theoretical curve eq. (\ref{TW1}), the points originate from 60000 numerically
generated WL matrices of size $N=80$ and $T=320$. To generate efficiently TW
densities, see \cite{edelmanNumerics}.
}
\end{figure*}
\end{center}

As an independent check we have generated real WL matrices and compared the
numerical distribution for the smallest and largest eigenvalues to $F_1(x)$,
see figure
\ref{TWnumcheck}.
The result for the largest and smallest eigenvalues given so far is valid for
the standard, Gaussian WL ensemble. Because of the implicit nature of the TW
distribution, we have not managed to analytically
derive the corresponding distribution
valid for the generalised ensemble (\ref{Pgen}) with power law tails.
In related models about the generalised WD class it was argued \cite{BBP,BCP} 
that a transition between TW and the Fr\'echet or 
normal distribution takes place for the largest eigenvalue.

\begin{center}
\begin{figure*}[h]
  \unitlength1.0cm
\centerline{  \epsfig{file=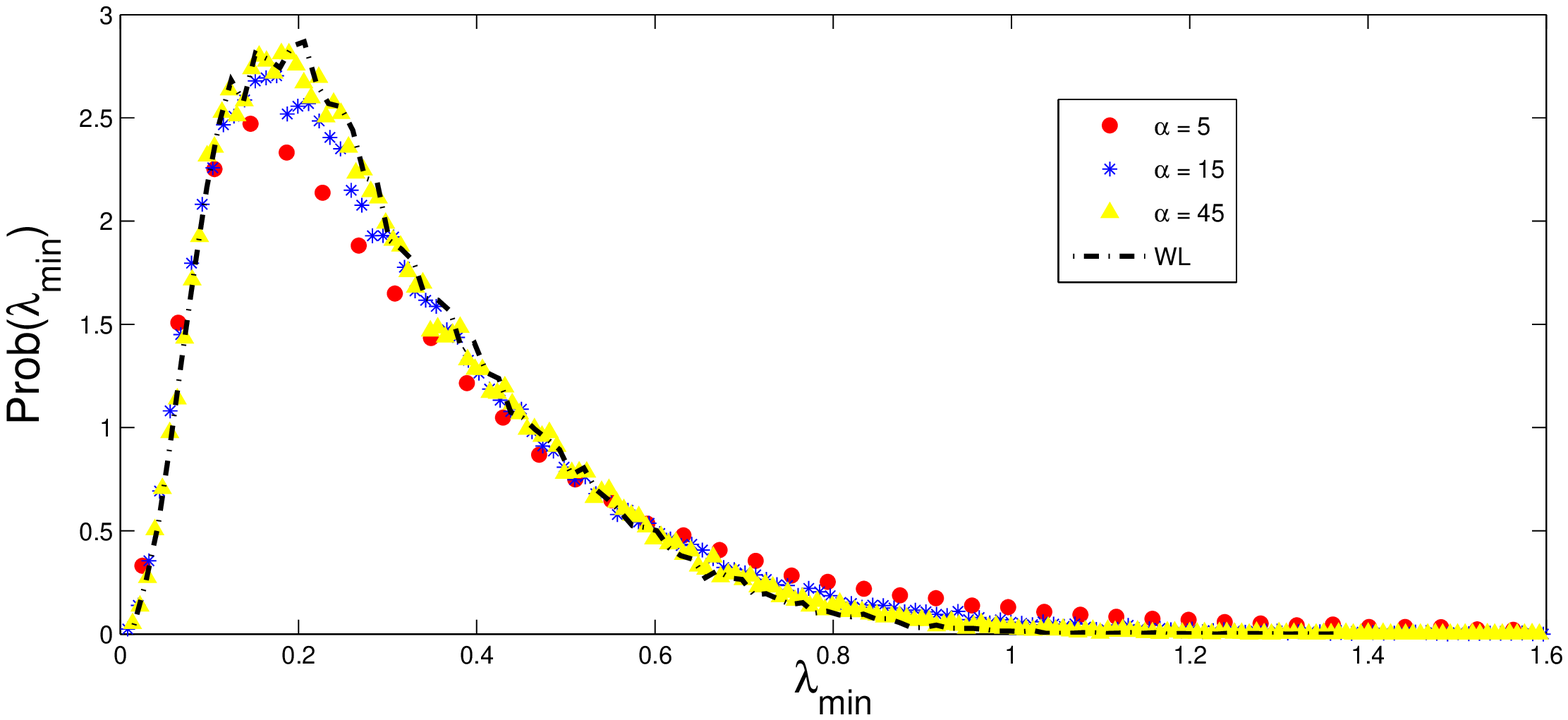,clip=,width=9.8cm}
\epsfig{file=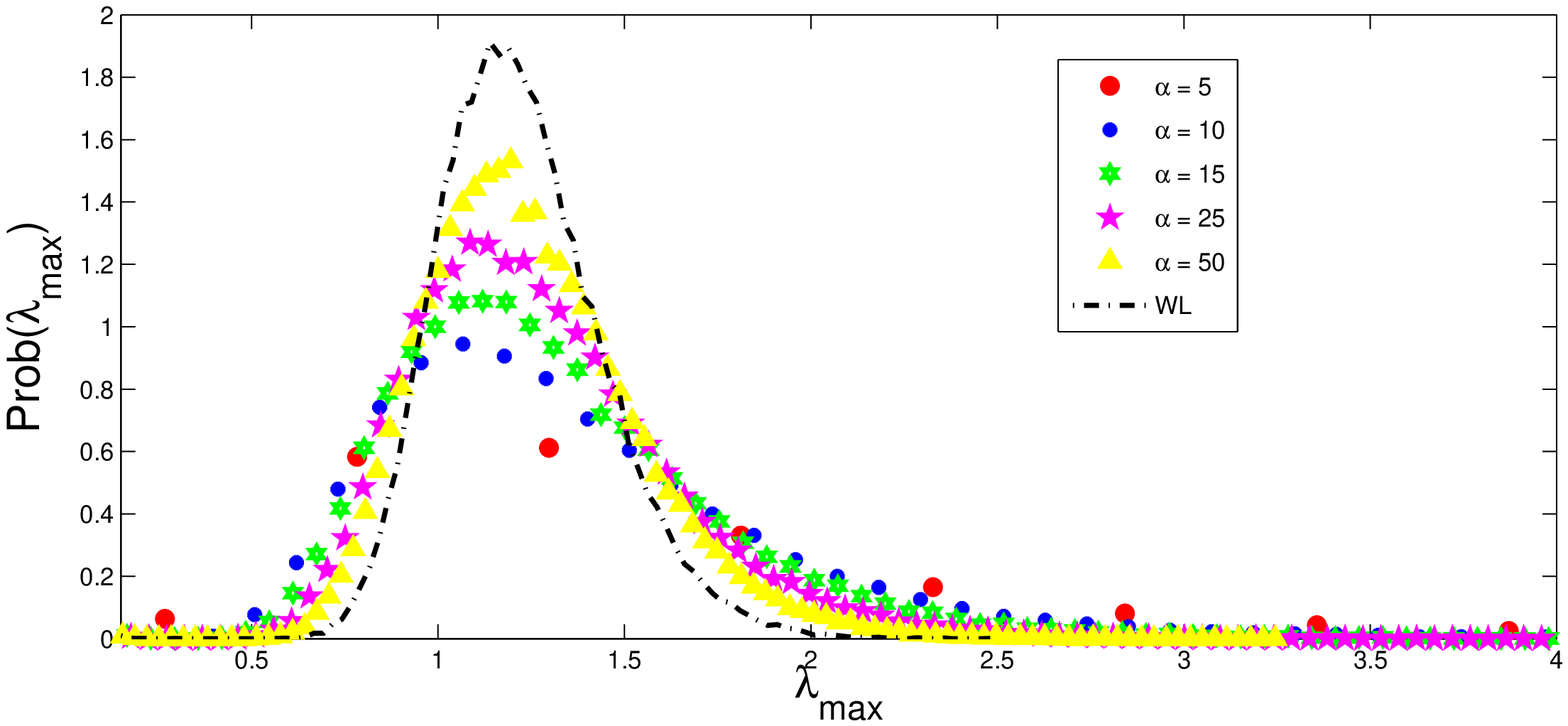,clip=,width=9.8cm}}
  \caption{Distribution of the {\it unscaled}
smallest (left) and largest (right) eigenvalues
  for the generalised
  ensemble eq. (\ref{Pgen}) for the case $N=8$, $T=15$ and different values of
  $\alpha$. For comparison
  the corresponding distribution for the WL ensemble (dash-dotted black line)
is given, which after a proper rescaling (eqs. (\ref{min}) and (\ref{max}))
maps to fig. \ref{TWnumcheck}.
}\label{TWgen}
\end{figure*}
\end{center}

Here, instead we have numerically produced the
distribution of the smallest and largest eigenvalue in the generalised
ensemble and observed the expected convergence to the WL distribution when
$\alpha$ gets large, see fig. \ref{TWgen}. Because we  do not know the
modification of the scaling relation in the generalised WL ensembles we show
the unscaled data.
An important observation we make is that the distribution of the largest
eigenvalue is strongly modified due to the power law tail of the spectrum. In
contrast the smallest eigenvalue undergoes only a mild modification. This is
probably because the exponential suppression of the spectral density at the
lower ``pseudo'' edge is not dramatically different from the square root edge
in WL.
This will be important when comparing to data in the next section.

\sect{Comparison to Financial Correlation Matrices}\label{data}

In this section, we compare the RMT predictions from the previous section
to data from financial covariance matrices.
Our motivation with respect to previous works has been two-fold.

First, can we improve
the fit of financial spectra to the global MP density of eigenvalues following
the null-hypothesis of Gaussian random variable by introducing correlations
among matrix elements that lead to power-law tails?
A similar route has been followed previously in \cite{biroli} albeit with
a different model; our model and a first comparison was given in \cite{AV}.

Second, can we test RMT predictions that are more robust under deformations of
the Gaussian measure of WL than the MP density? The quantities we have looked
at are the \emph{individual} and \emph{global} spacing distribution for
standard WL, and the individual distribution of the smallest
eigenvalue. The individual distributions of course require the definition of a
suitable \emph{ensemble} of homogeneous matrix objects to average over.
The global spacing distribution was previously compared to data in
\cite{plerouplus,Plerouetal}.

In the first subsection \ref{global},
we focus on quantities that can be inferred from a
\emph{single} financial covariance matrix, whereas in the second subsection
\ref{indiv} we define a new \emph{chopping procedure} to generate {\it
  ensembles} of covariance matrices from one single instance in order to
analyse the individual eigenvalue statistics.

\subsection{Power-law decay in the global density and the global spacing}
\label{global}

In this subsection we have analysed a single set of data given by $N=401$
stocks and a time series of daily prices $T=970$.
The data were obtained from the S\&P 500 Equity Index. It is representative
of the U.S. equity market and hence of their economy including the 500 largest
companies. We extracted the daily opening prices for the 4-year period
2002-2006. Stocks that did not survive in the index during that period were
deleted from the analysis, hence the reduced number of 401 stocks.
Henceforth we will refer to
this dataset as set SP.

\begin{center}
\begin{figure}[h]
\centerline{\epsfig{file=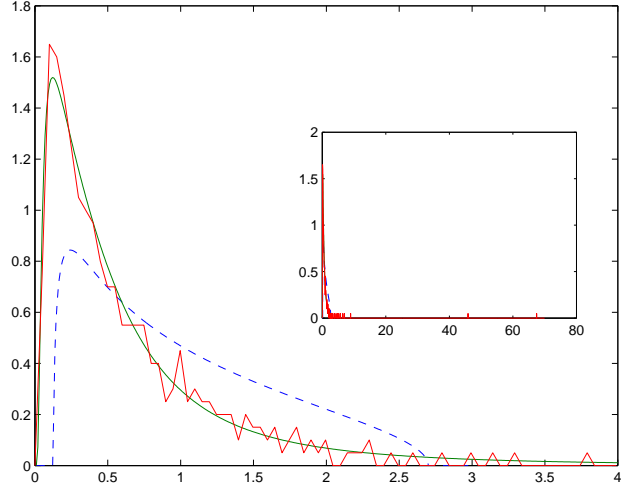,clip=,width=10cm}}
\caption{Comparison between part of the rescaled eigenvalue distribution from
  financial data set SP,
and the macroscopic density $\rho_\al(x)$ from RMT
for the generalised model eq.
  \eqref{rhogen}, in red and green respectively.
The best fit gives a value of $\al\approx 0.95$, which corresponds
to a power-law decay as $\rho_\al(x)\sim x^{-2.95}$.
For comparison we have added the parameter free MP density eq. (\ref{MP}).
The inset shows all eigenvalues on a different scale.} \label{Comparison}
\end{figure}
\end{center}

The financial covariance matrix is obtained in the standard way \cite{laloux}
which we briefly recall.
First the normalised return $G_i(t)$ of the $i$-th stock at time $t$ is
computed as the following function of its stock price $S_i(t)$
\be
G_i(t) =\log[S_i(t+\Delta t)]-\log[S_i(t)] \ ,
\label{return}
\ee
where the time step $\Delta t$ is 1 day for our set SP.
Then the temporal mean $\langle G_i(t)\rangle$ and variance $\sigma_i$ are
computed for each stock independently to give the normalised data matrix
elements
\be
X_{it}=  \frac{G_i(t)\ -\ \langle G_i(t)\rangle}{\sigma_i}\ .
\label{fincov}
\ee
Clearly the $X_{it}$ over the period $t=1,\ldots,T$ have variance one and
vanishing mean. Then the eigenvalues of the covariance matrix
$\mathbf{C}=(1/T)\mathbf{X}^{T}\mathbf{X}$ of these data are computed
and histogrammed in fig \ref{Comparison}.
The spectral density obtained in this fashion is normalised to unity and
rescaled to have a first moment equal to $1$.  It can then be compared to the
MP density (\ref{MP}) which is parameter free under the same rescaling, and
to the generalised density in \eqref{rhogen}.
In previous works using the standard Gaussian
WL ensemble a one-parameter fit was performed using a variance dependent
MP density, arguing that
the outlying eigenvalues with respect to RMT would effectively reduce the
inverse variance 
$\sigma$ in eq. (\ref{PWL}) \cite{laloux}.
No such freedom is left in our approach as the first moment has
been already normalised to $1$.

The resulting agreement between the generalised density \eqref{rhogen} and the
data is very good after fitting the only free parameter $\al\approx 0.95...$
in the generalised model. This leads to a power-law decay with exponent
$\al+2\approx 2.95$. It has been
observed independently in the framework of a similar model \cite{biroli}
that the global spectral histogram is better fitted by a
density with power-law tails rather than MP. Despite many differences between
the two models, the exponent $2.93$ found in
\cite{biroli} is similar to ours, using a very similar set of data
(daily returns of from S\&P 500 Equity Index from 2003-2007).

\ \\

A second quantity we analyse is the nearest-neighbour spacing distribution
which resolves local information of the order $O(1/N)$ and is much more
universal, being the
Fredholm determinant of the universal sine-kernel \cite{Mehta}. For a detailed
discussion of universality issues we
refer to \cite{AV}.

In this subsection we specifically look at
the \emph{global} level spacing distribution, obtained from
a \emph{single} sequence of sorted eigenvalues (one matrix sample).

\begin{center}
\begin{figure}[t]
\centerline{
\epsfig{file=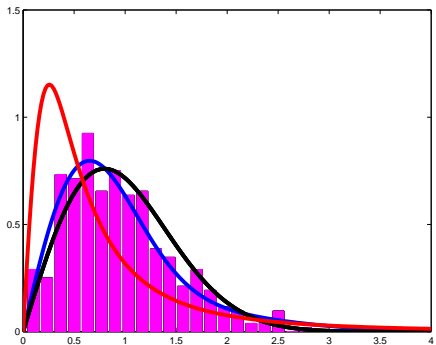,
clip=,width=7.0cm}}
\caption{The unfolded global spacing distribution from
  financial data set SP vs the RMT prediction
eq. (\ref{WD}) for WL (black right curve), and eq. (\ref{genspace})
for our generalised ensemble: $\al=0.95$ (red left curve) and the best fit
with  $\al=8$ (blue middle curve).
} \label{globalspace}
\end{figure}
\end{center}

After unfolding the spectrum, a standard procedure in RMT \cite{GMW} where we
use a polynomial fitting of the cumulative eigenvalue distribution
(see appendix \ref{unfold} for details),
the normalised histogram of \emph{all}
spacings between consecutive eigenvalues is plotted in
fig. \ref{globalspace}.
Note that we have made an important assumption here: we assumed
that averaging over the spacings between different consecutive eigenvalues
of a single covariance matrix converges to a single distribution function
after unfolding.
Whereas in RMT this is known to hold due to the self-averaging property (or ergodicity)
this is by no means guaranteed for financial covariance matrices.

The result in fig. \ref{globalspace} shows that the spacing distribution
seems to deviate from the WS of standard WL, although our statistics is not
extremely good. This is in contrast with previous analysis
\cite{plerouplus,Plerouetal} where a good agreement with WS was found
\footnote{In the same paper no significant deviation from MP
was found either, in contrast to our fig. \ref{Comparison}.}.
For comparison we give our generalised spacing
distribution eq. (\ref{genspace}): from the value $\al=0.95$ as determined
in fig. \ref{Comparison} we get a worse fit than from the WS. However, using
$\alpha$ as a free fitting parameter the apparently best fit is given by the
generalised spacing for  $\al\approx8$, thus parametrising the deviation from
WL. This can be interpreted as the system not being fully random, as we have
seen from the global spectrum.

Because our statistics is not very good we will come back to this question at
the end of the next subsection \ref{indiv}. The ensembles from a
chopped covariance matrix show the same effect more clearly.

\subsection{Universal individual eigenvalue distributions}
\label{indiv}

In this section we analyse the distribution of individual eigenvalues and
spacings and compare them to the universal TW and WS distributions
respectively (and their generalisation). In order
to generate ensembles of covariance matrices we use two different methods here:
\\

{\it Method 1:} We take a long time series $T$ of
a small number $N$ of stocks, and then chop $T$ into $\ell$ smaller
time windows of equal length $t$, $T=\ell t$. This creates an ensemble of
$\ell$
covariance matrices of size $t\times N$ of the {\it same} $N$ stocks,
with $c=N/t\in(0,1]$.\\

{\it Method 2:} We take a time series $T$ of $N$ stocks, and chop both
$N$ into $k$ sets of $n$ stocks, with $N=kn$, as well as chopping
 $T$ into $\ell$ smaller time windows of equal length $t$, $T=\ell t$, as
before. This creates an ensemble of $k\ell$
covariance matrices of size $t\times n$, thus {\it mixing different} sets
of $n$ stocks,
with $c=n/t\in(0,1]$.\\

While the chopping in smaller time windows in method 1 (and 2) is unambiguous
- apart from choosing the length of the time window $t$ -
the chopping into smaller subsets of stocks
may seem at first quite hazardous, mixing in an arbitrary way
fairly heterogeneous data.
Surprisingly, we find the same \emph{universal} properties from both methods.

To test the consistency of method 2 we have produced ensembles of the same
size using several choppings and including different subsets of stocks (see
appendix \ref{chopperm}), and we detected a substantial robustness of ensemble
properties with respect to random reshuffling.
A clear advantage of method 2 is that we can produce a much better
statistics.
Furthermore we have varied the length on the time window $t$ in both method 1
and 2. Here we have taken care to choose $t$ (and $n$) such
that the resulting value $1/c$ remains bounded of the order $O(10)$, in order
to avoid the problems discussed in \cite{DKO} for large ratios.

We begin by using ensembles generated from method 1. This is done with our
second data set DJ based on the Dow-Jones Index obtained as
follows. Daily returns for $N=27$ stocks were used in the time interval from
June 1997 to October 2005, leading to a $T=17109$. From this we have created
an ensemble of 179 (356) correlation matrices of size $96\times 20$ ($48\times
20$). Our choice was motivated by having some data with
the same $c$-value as from method 2 below.
\begin{center}
\begin{figure}[htb]
\centerline{
\epsfig{file=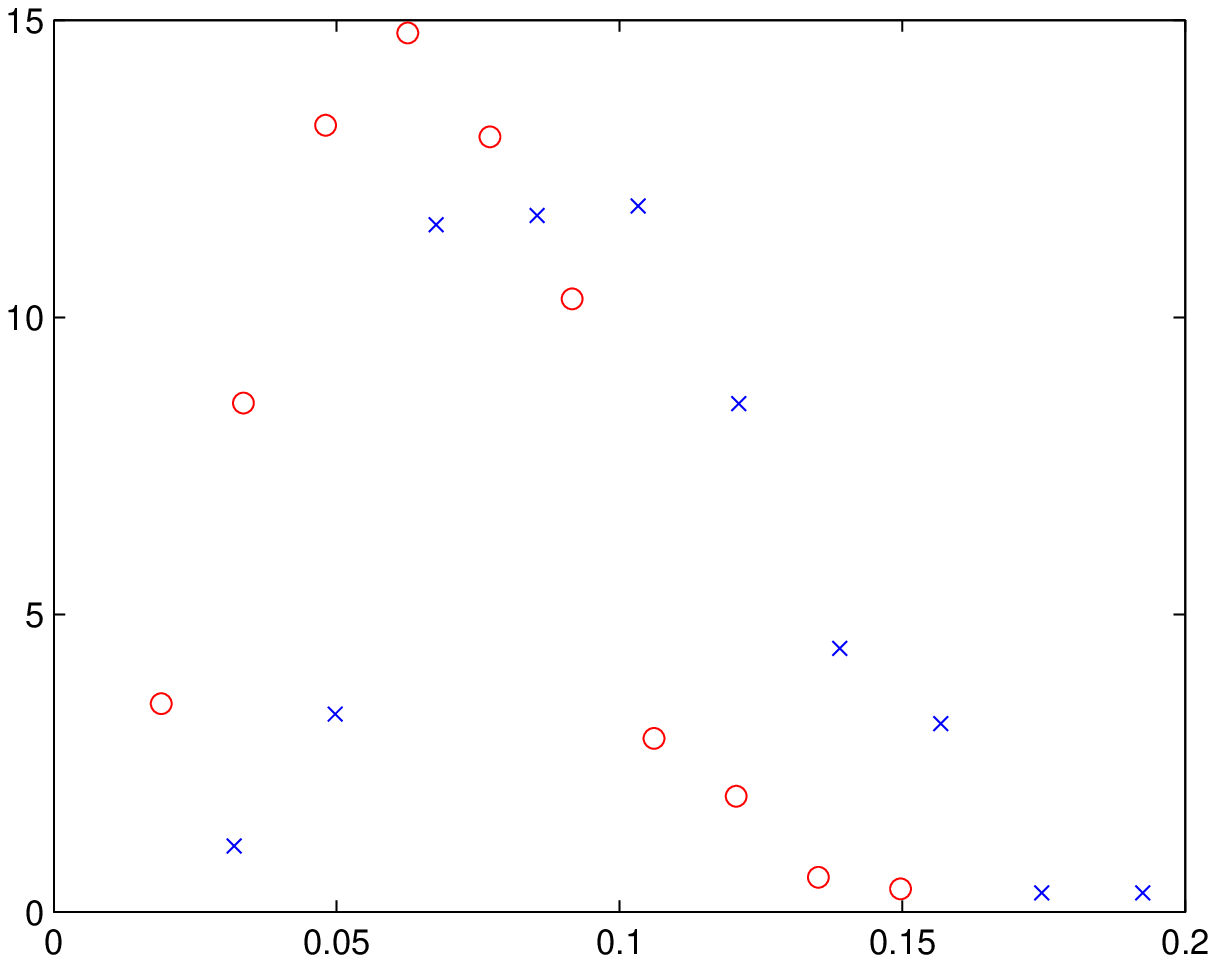,clip=,width=5.8cm}
\epsfig{file=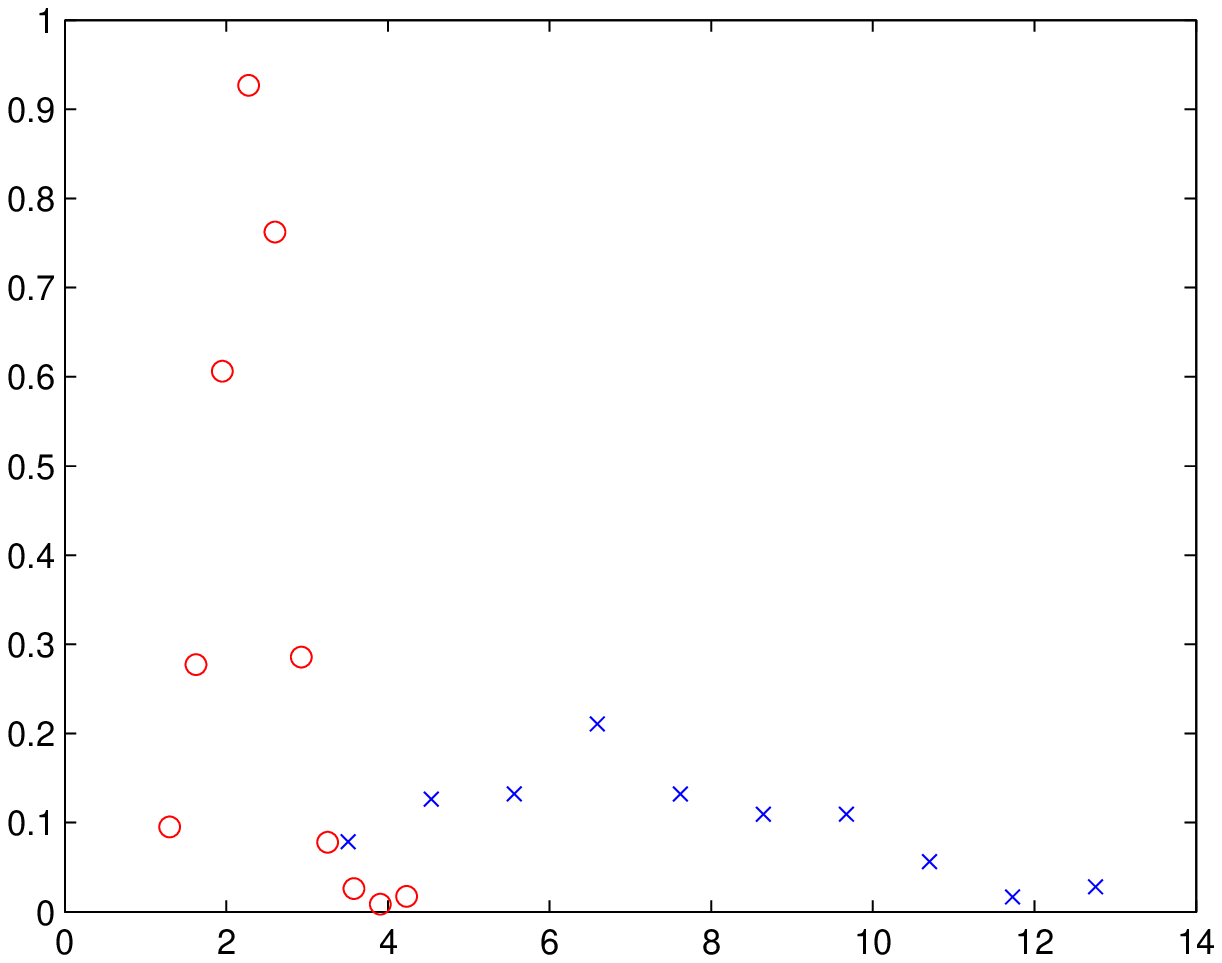,clip=,width=5.8cm}
}
\caption{Method 1 using data set DJ and the ensemble
$48\times20$. Left plot: superposition of the smallest (red circles) and
second smallest eigenvalue (blue crosses).
Right plot: superposition of the second largest (red circles) and largest
eigenvalues (blue crosses).
} \label{rawsmallR}
\end{figure}
\end{center}
The full spectrum is given in fig. \ref{DJ2048rho} in appendix
\ref{macrochop}, but we shall focus on the local properties here.
The result for the unscaled
first and second
eigenvalue is given in fig. \ref{rawsmallR} left. The spacing
between the two (as well as their individual widths)
are of the order of the mean level spacing
$1/N\approx 0.05$. This indicates that a comparison to the TW
prediction from RMT (and its generalisation) is meaningful.

For comparison we also give the distribution of the largest and second largest
eigenvalue from 
method 1 in fig. \ref{rawsmallR} right. Obviously the shape and width are very
different, and the spacing is much larger than $1/N$. Already the second
largest 
eigenvalue is within the support of both the MP and generalised density, see
fig. \ref{DJ2048rho} of the full spectrum in appendix \ref{macrochop}.
From previous considerations  \cite{laloux}
as well as our previous section \ref{data} one could speculate
whether or not the largest eigenvalues are non-random, containing
information.

Because in the WL ensembles TW describes both the smallest and the largest
eigenvalues we could try to compare these predictions with the data.
However, there are some hints that such a program is essentially ill-defined,
due to the absence of a clear-cut separation between eigenvalues following RMT
statistics and those carrying genuine information. The fact that we find a fit
better fit to our generalised RMT with power law tails makes this even more
difficult.
On the other hand it has been argued in \cite{GK} that information could be
extracted from underneath the density close to the right edge, using a power
mapping method.
\begin{center}
\begin{figure}[htb]
\centerline{
\epsfig{file=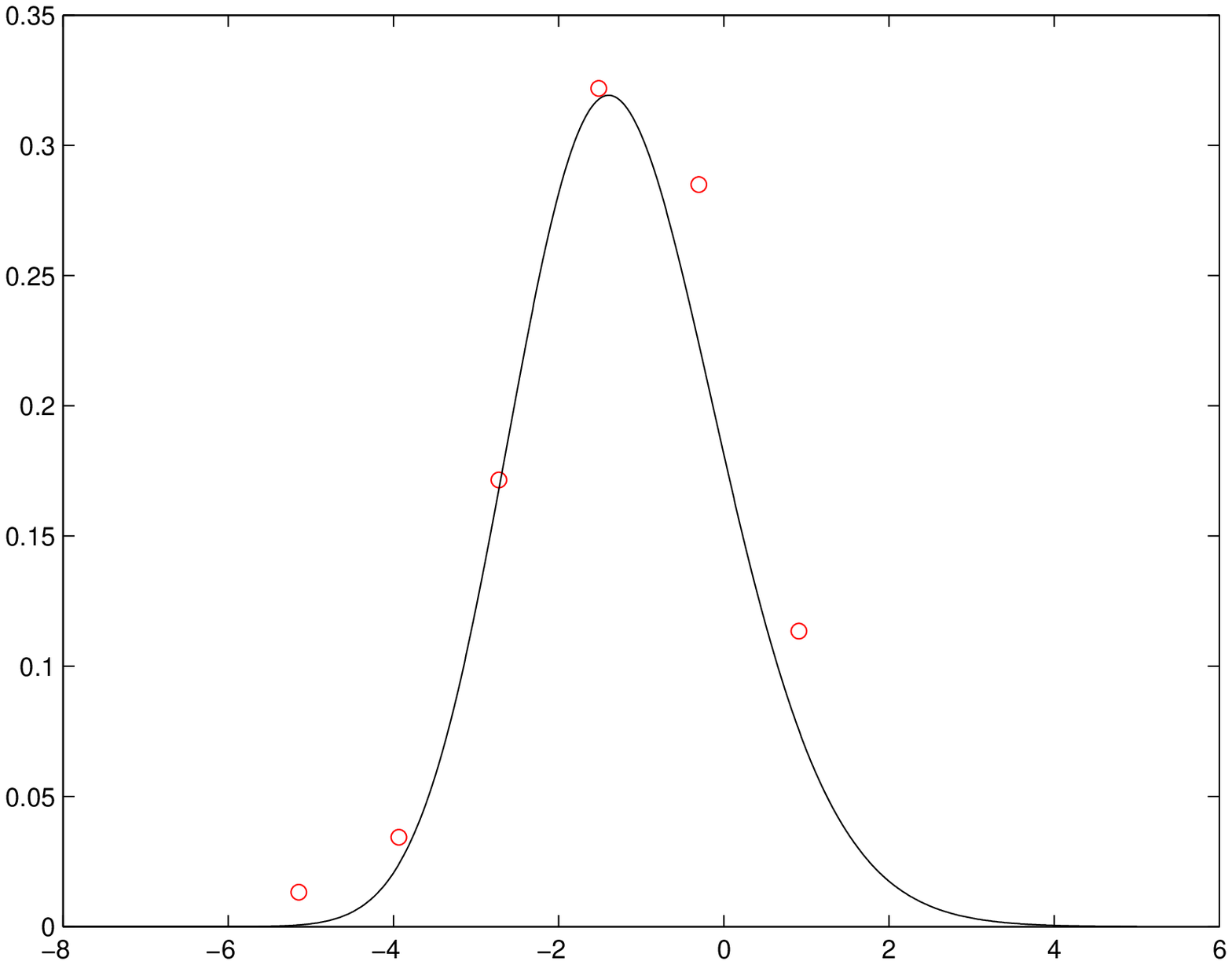,clip=,width=5.8cm}
\epsfig{file=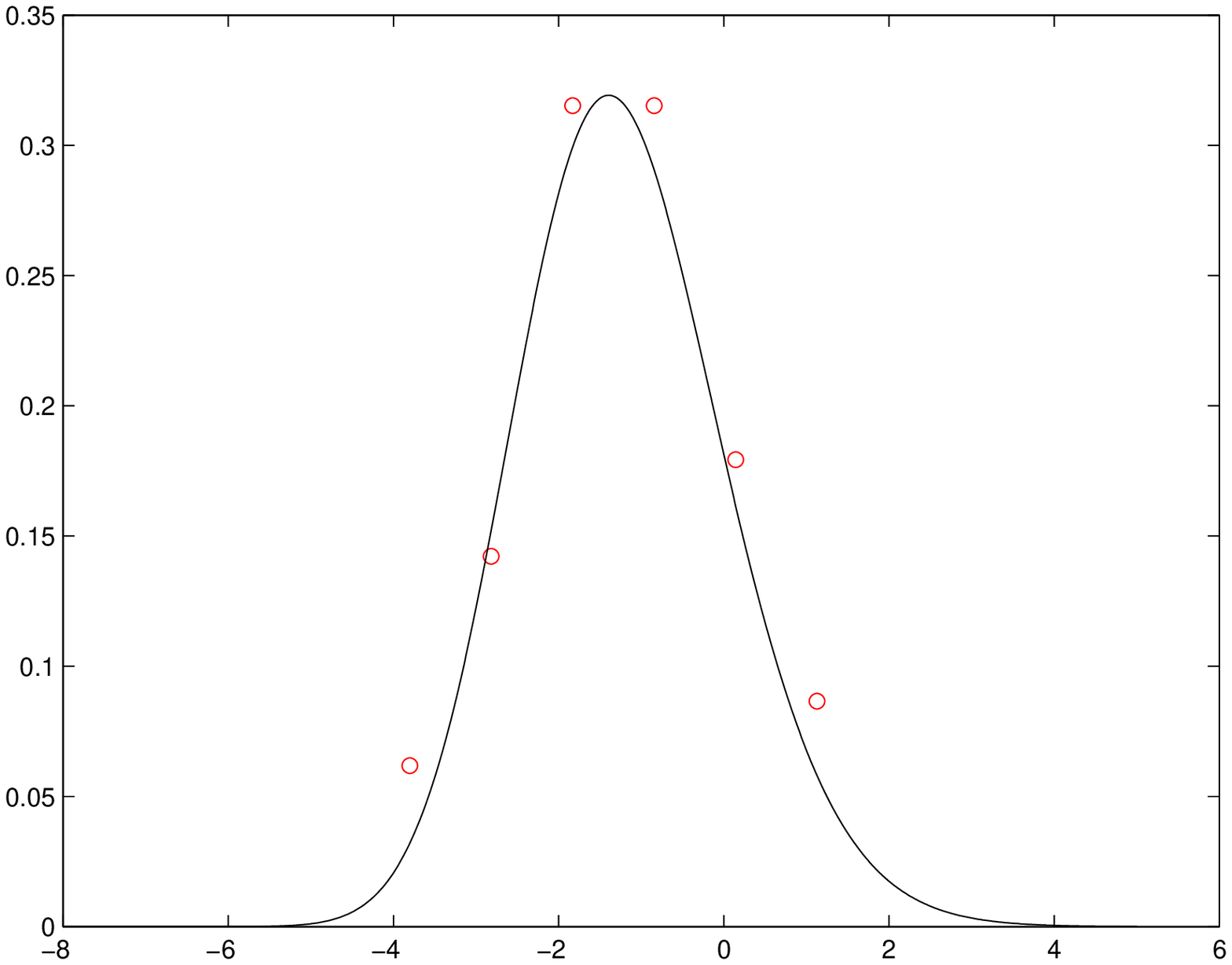,clip=,width=5.8cm}
}
\caption{The rescaled smallest eigenvalue from method 1 using data set DJ.
Left: ensemble of size $48\times 20$. Right: ensemble of size $96\times 20$.
} \label{TWfit}
\end{figure}
\end{center}

For these difficulties
we have fitted only the smallest eigenvalue with the TW distribution, with
the result given in fig. \ref{TWfit}. As was discussed in the previous
subsection  \ref{global} we only expect minor deviations for the smallest
eigenvalue in our generalised ensemble eq. (\ref{Pgen}).
Although we only have a few points in the plot, we can at least conclude that
the data are consistent with at fit to TW (as well as our generalised model).
Note that this finding is invariant under doubling the size of the temporal
windows, fig. \ref{TWfit} left vs right.

The TW distribution is well know to be strongly universal within RMT, being
invariant under non-Gaussian deformations of the probability distribution
eq. (\ref{PWL}), by adding higher order terms in the exponent. We have
therefore established that a trace of this 
universality is also seen in ensembles of financial covariance matrices.

\begin{center}
\begin{figure}[b]
\centerline{
\epsfig{file=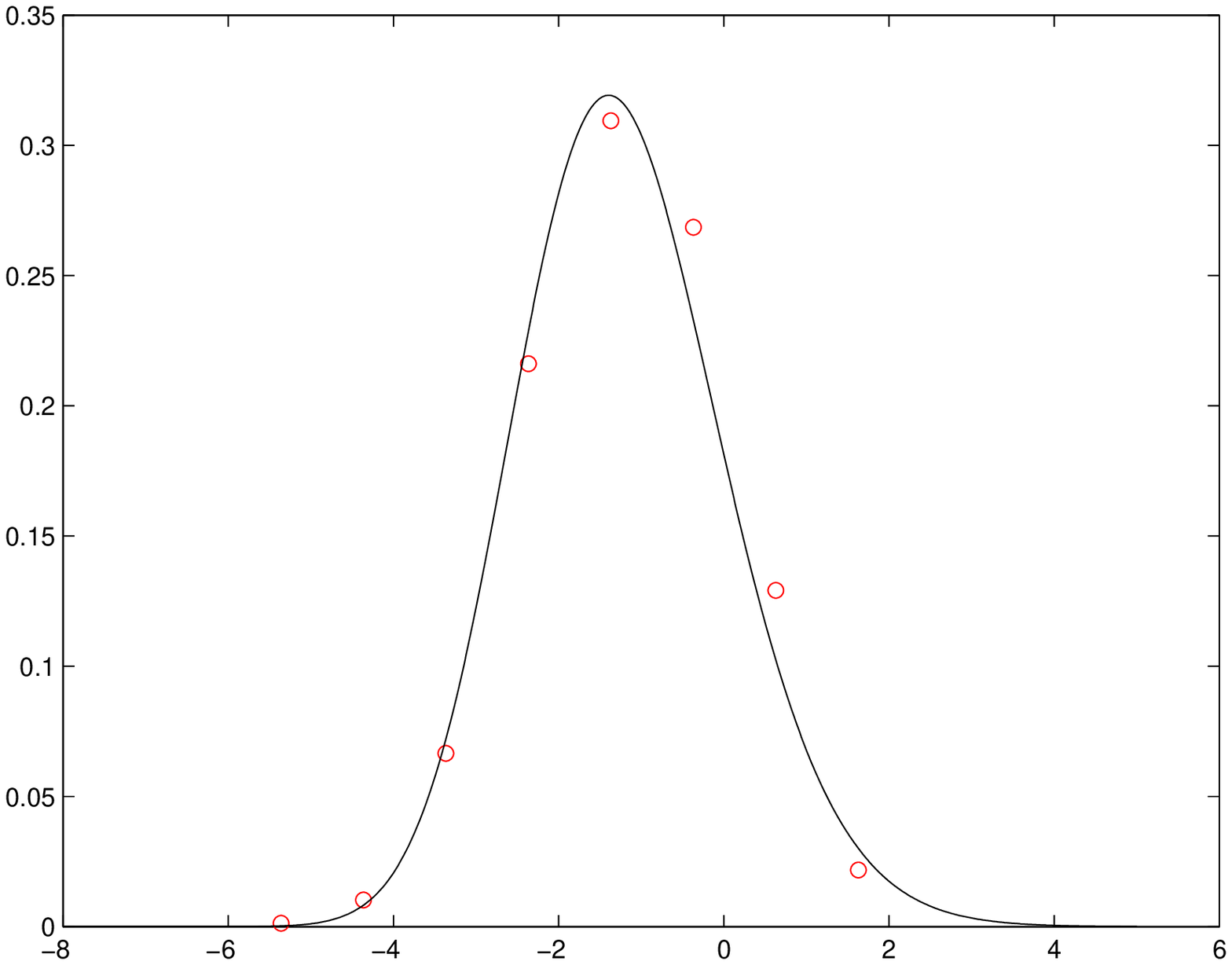,clip=,width=5.8cm}
\epsfig{file=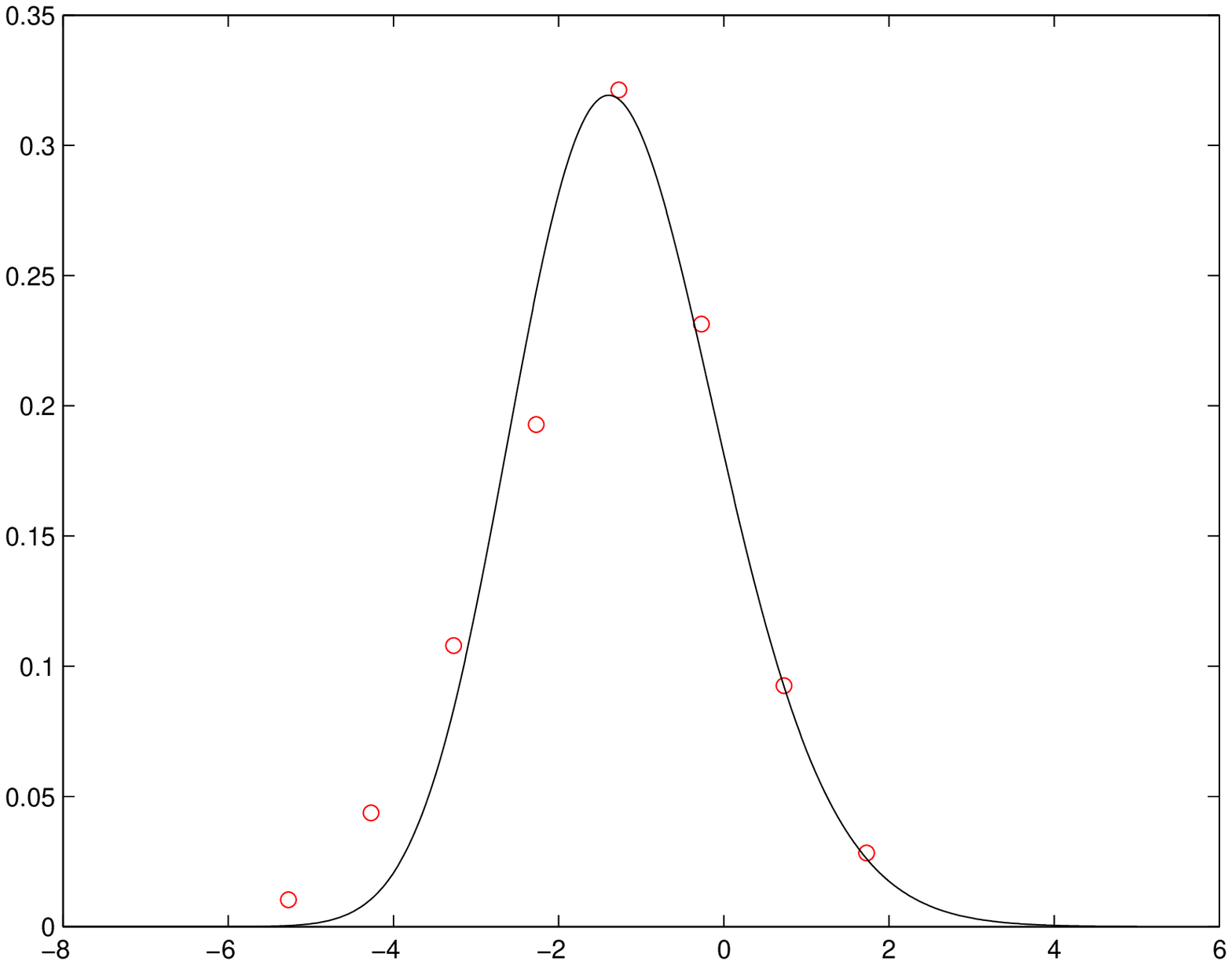,clip=,width=5.8cm}
\epsfig{file=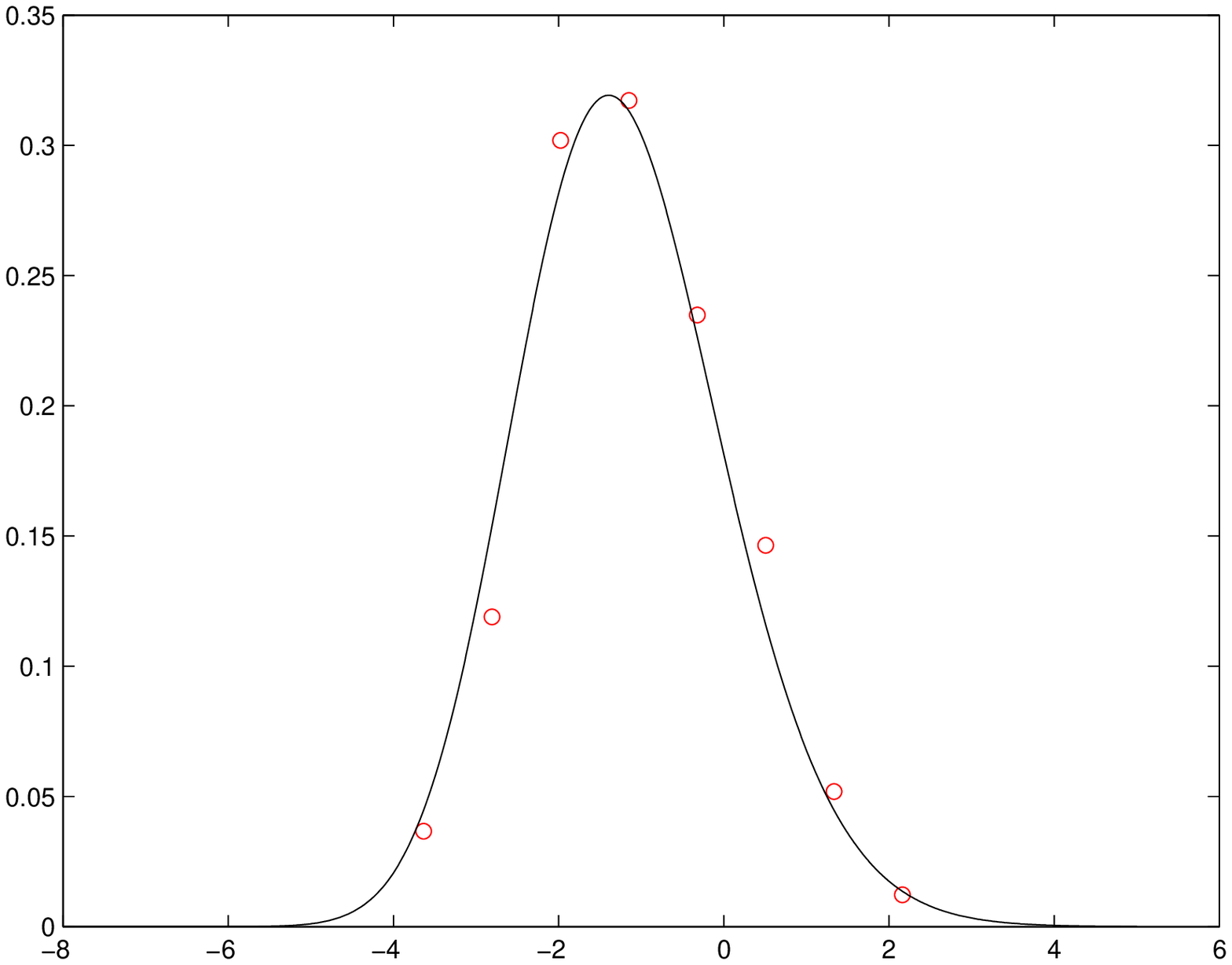,clip=,width=5.8cm}
}
\caption{The rescaled smallest eigenvalue from method 2 using data set SP.
Left: ensemble of size $48\times 10$, middle $48\times 20$, right:
$96\times 20$.
} \label{TWfit2}
\end{figure}
\end{center}
We can now repeat the same analysis using method 2. Here we reuse the data set
SP from the previous subsection \ref{global}, and chop the $970\times401$
matrix into 
ensembles of submatrices of various sizes. The full densities for submatrix
size $48\times10$ and $48\times 20$ of respective ensemble sizes 800 and 400,
are shown in appendix \ref{macrochop} fig. \ref{SPrho}. The situation for the
distribution of the first vs second, as well as largest vs second largest
eigenvalue is similar as in method 1 described above, and we refer to appendix
\ref{chopperm} for details. The fit to the universal TW distribution using
method 2 is shown in fig. \ref{TWfit2}, with the same conclusion as above.
Here we have halved the chopping size in $n$ (stock) direction, as well as
doubled the time window $t$, which all lead to consistent fits.

In appendix \ref{chopperm} we also compared the distribution of the smallest
eigenvalues for a fixed sub-matrix size $48\times 20$, choosing different
subsets of 20 stocks out of 400. Because the unscaled first
eigenvalues lie on top of each other we do not repeat the rescaled fit to TW
for these permutations. We have also analysed other sub-matrix sizes, and we
have 
subpartitioned the data set DJ using method 2 as well. In all cases we obtained
approximately the same level of consistency.\\

We now move to the second local quantity,
the individual spacing distributions for our ensembles
as described in eq. (\ref{individual}). Because of better statistics we
restrict ourselves to ensembles generated from method 2. Since we are dealing
with individual spacings, no unfolding is necessary here.
All distributions are consistent with the WS eq. (\ref{WD}) relevant for the
unperturbed WL ensembles. We do not find any trace here from our
generalised ensemble, in spite of its better fit of the full density, see
fig. \ref{SPrho}.
 \begin{center}
\begin{figure}[htb]
\centerline{
\epsfig{file=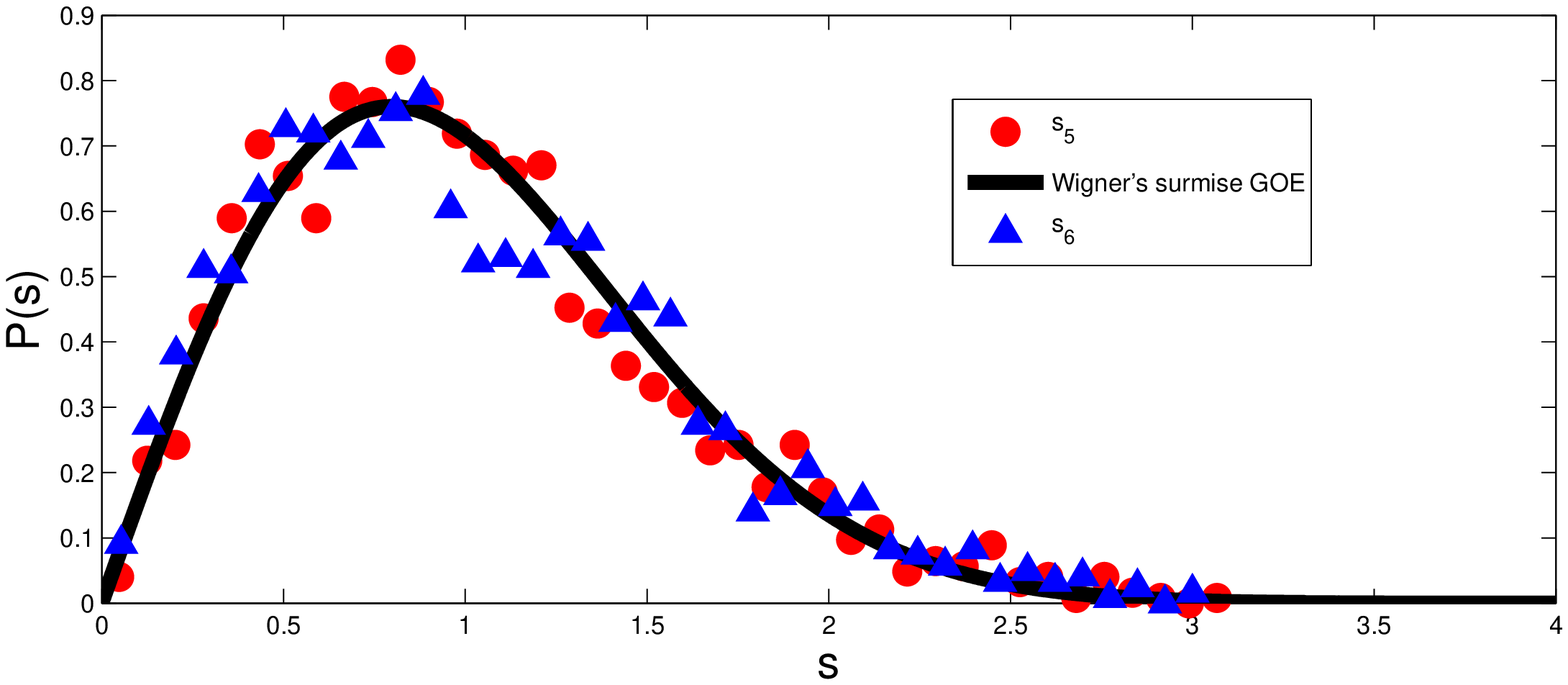,clip=,width=9.8cm}
\epsfig{file=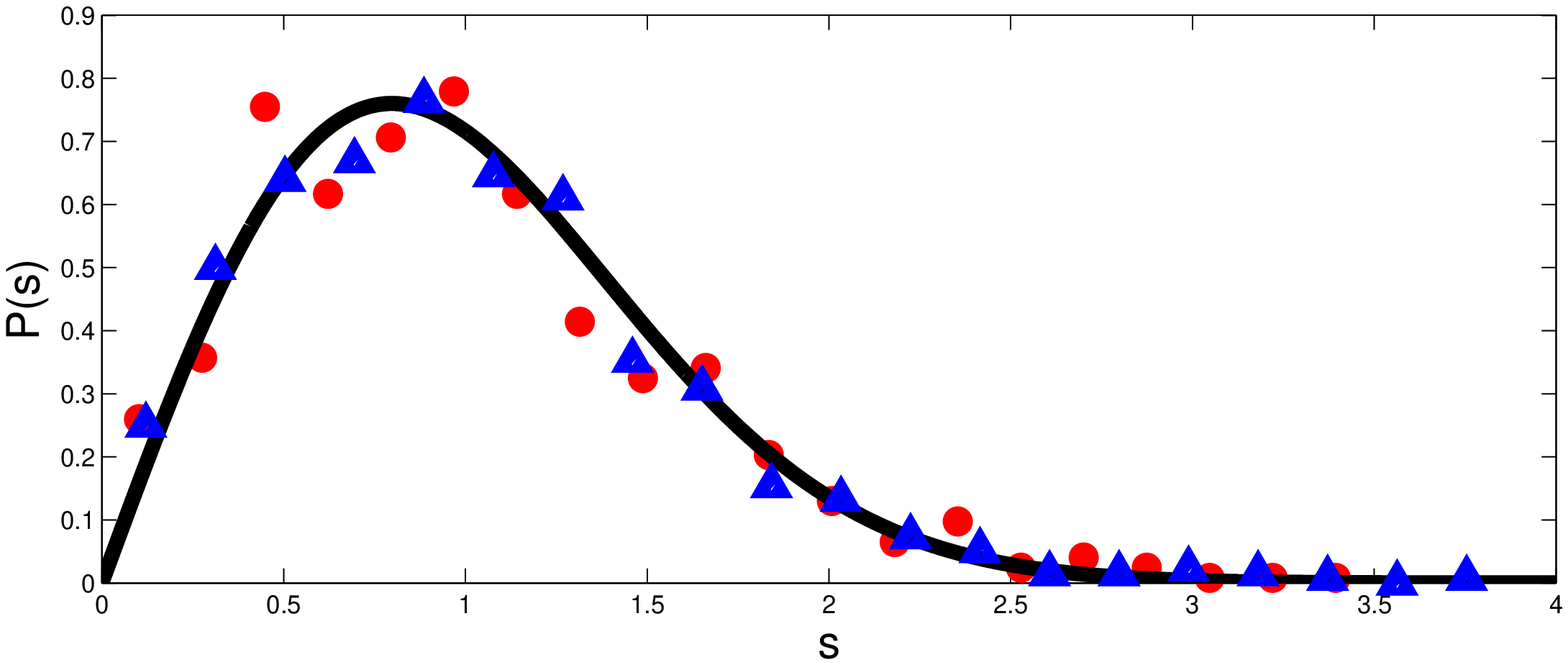,clip=,width=9.8cm}
}
\caption{The WS eq. (\ref{WD}) vs the
distribution of two individual spacings in the middle of the
  spectrum: 5th and 6th spacing of 10 eigenvalues, for ensembles of size
$24\times 10$ (left) and $48\times 10$ (right) from method 2.
} \label{indivspace}
\end{figure}
\end{center}

\begin{center}
\begin{figure}[bp!]
\centerline{
\epsfig{file=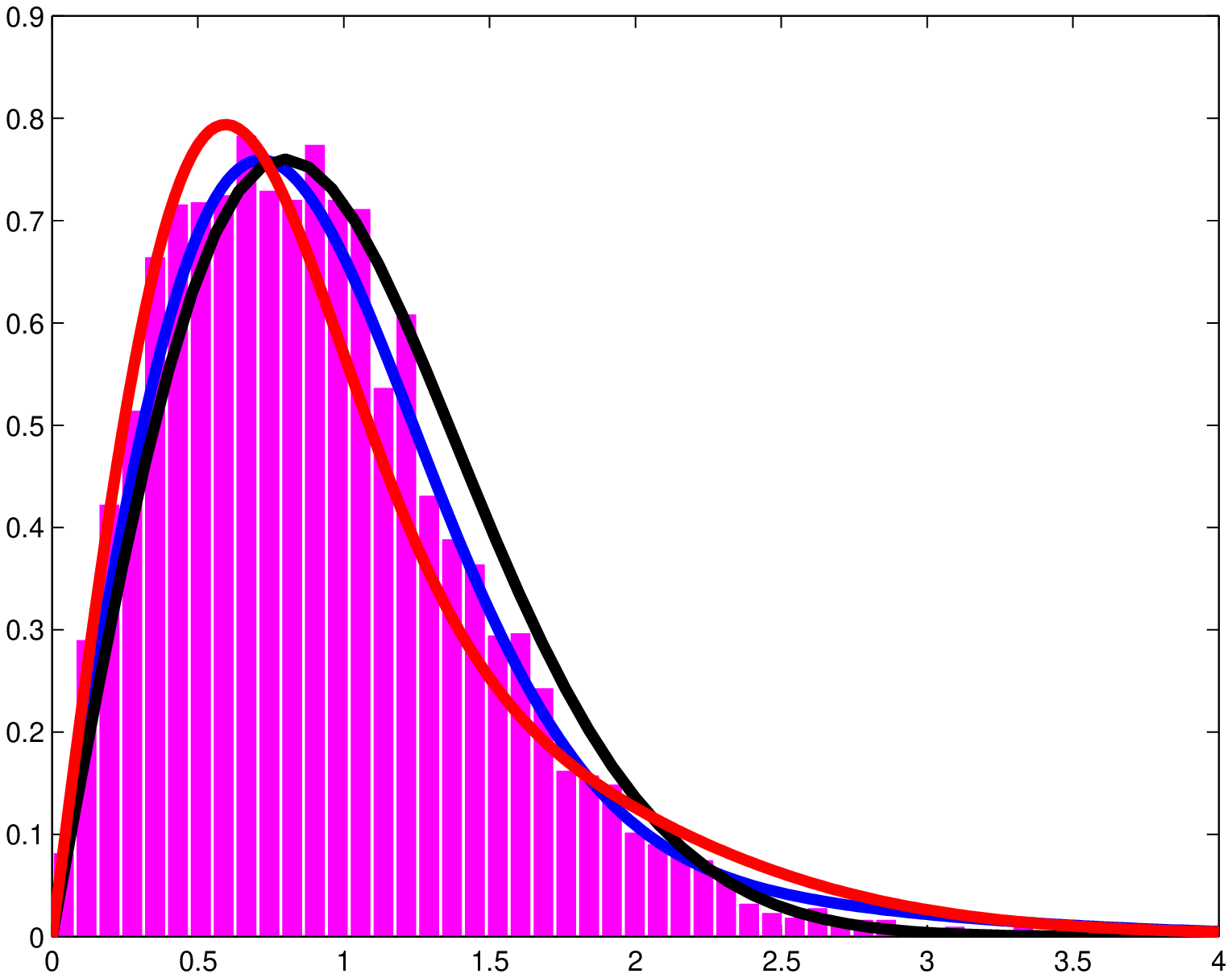,
clip=,width=7.8cm}
\epsfig{file=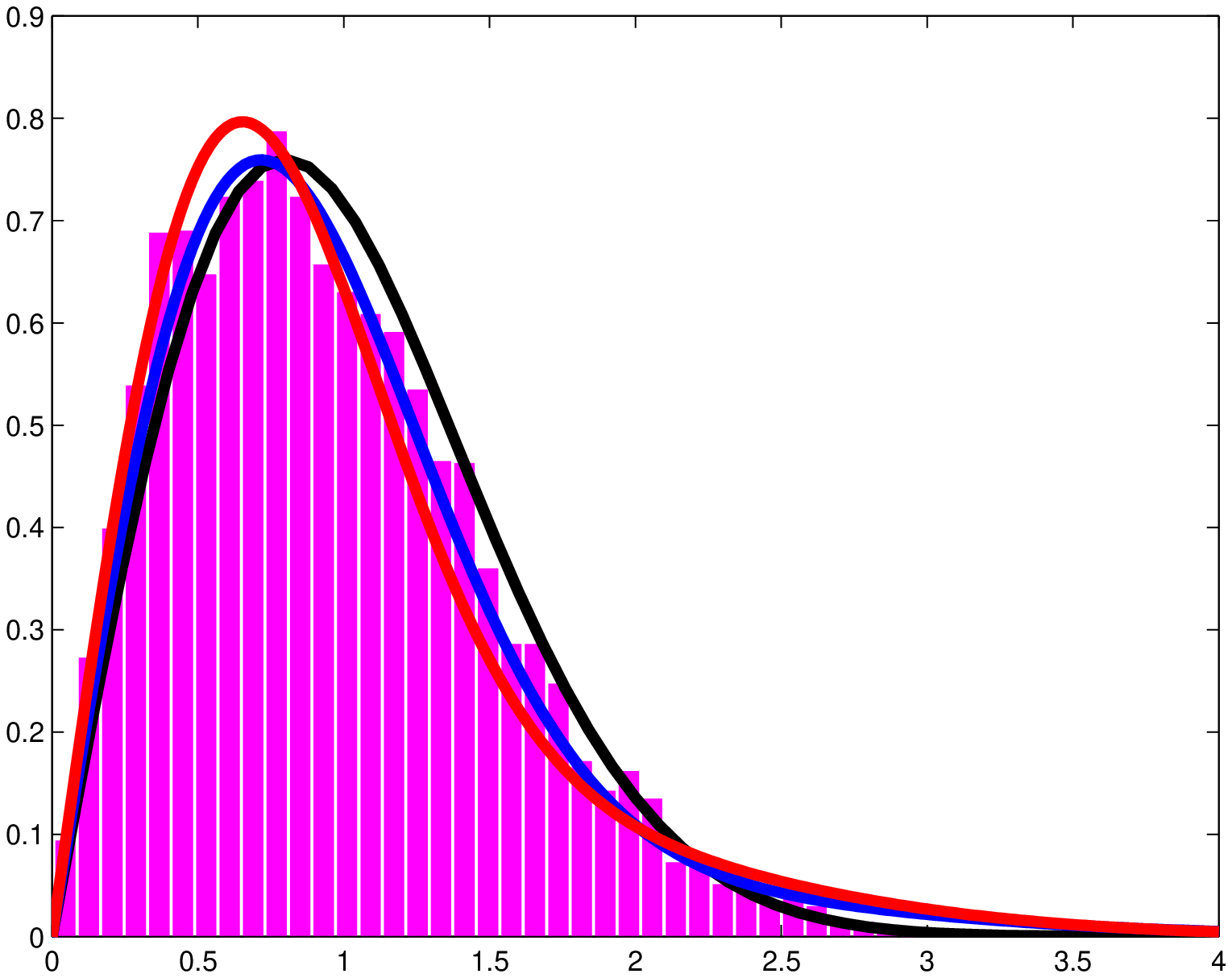,
clip=,width=7.8cm}
}
\caption{The WS eq. (\ref{WD}) vs the
global spacings distribution, for ensembles of size
$48\times 10$ (left) and $48\times 20$ (right) from method 2.
The generalised surmise is displayed with the values $\alpha=7,8$
extracted from the global density in fig. \ref{SPrho} respectively, and with
the best fitted 
value of $\al=9$ in both cases.
} \label{globalspace2}
\end{figure}
\end{center}
Let us return to the question of a deviation from the WS for the global
spacing distribution from the previous subsection \ref{global}. Repeating the
same analysis by unfolding the chopped data from method 2 and then averaging
over all consecutive spacings we again find a deviation from the WS in the
global spacing distribution, but with a much clearer signal compared to fig.
\ref{globalspace}.

As in fig.  \ref{globalspace} we compare the WS and the generalised spacing
eq. (\ref{genspace}). Like before the $\alpha$-value determined from the global
spectral density, see fig. \ref{SPrho}, does not give the
best fit (although it is not as bad as in fig. \ref{globalspace}). Using
$\al$ as a free fit parameter
we can clearly give a much better fit than the WS. This may again be seen as an
indication that the data are not fully random.

\sect{Conclusions}\label{con}

We have studied the spectral properties of empirical covariance matrices from
financial data, by comparing to the predictions of two different Random Matrix
ensembles with uncorrelated and power-law correlated random variables.
While previous works have mainly focused on global properties of the
spectrum we were interested in
local properties such as the distribution of individual
eigenvalues or the individual spacings between eigenvalues.
The reason for investigating this matter was the strong degree of universality
of these quantities within RMT. These individual quantities can be looked at
without assuming self-averaging or ergodicity, which is {\it apriori} true only
within RMT.

This question automatically drove us to define ensembles of empirical
covariance matrices starting from a fixed set of time series of different
stock prices. Two methods were devised to 
generate such ensembles, either by
averaging over just different time windows, or also over
different sets of stocks. Within both approaches we found that our results
for individual quantities are
compatible with the standard Wishart-Laguerre (WL) ensembles of RMT,
starting from the null-hypothesis of Gaussian
random variables for the matrix elements;
the distribution of the smallest eigenvalue in our ensembles upon centring
and rescaling agrees with the
universal Tracy-Widom distribution, and the spacing between the $k$th and
$(k+1)$st eigenvalue for a fixed $k$ in the bulk of the spectrum follows the
Wigner Surmise (WS).

The reason for comparing with two different ensembles was motivated by previous
findings for the global spectrum; its fit to the Mar\v{c}enko-Pastur (MP)
density from Gaussian WL, which is known to be only weakly universal, could be
improved introducing correlations among matrix elements that lead to a
power-law decay. Such a deformed RMT was introduced by several authors
and can be based on a deformed entropy or superstatistical approach,
tailored to describe non-equilibrium or not fully random problems.

The question was whether or not this power-law determined from an excellent
one-parameter fit to the global density would also show up in the local
statistics of the data. While we failed to detect this in the individual
distributions - we expected only a small deformation for the smallest
eigenvalue - we did see such a deviation in the global spacing distribution,
where the average is taken over all consecutive spacings after unfolding.
This was at least qualitatively the case for
both a single covariance matrix and
the ensemble of such matrices produced from method 2.
The corresponding generalised WS was derived and numerically tested
within the generalised RMT ensemble with power-law tails.

The appearance of power-laws in the global spectrum on one hand,
and more robust universal local RMT correlations on the other hand
are quite reminiscent of findings in complex
networks. Also here first investigations gave an agreement with the WS
and  spectral rigidity \cite{jalan1}, whereas recent results for the
latter follow a generalised RMT \cite{jalan2}.

We conclude by mentioning some open problems.
The distribution of individual eigenvalues in the bulk is also known in
principle. However, its implicit, almost Gaussian form together with our
limited statistics makes it harder to detect than the skewed TW distribution.

Finally we could ask ourselves how much the chopping procedures introduced to
study individual properties wash out of the relevant
information. Clearly the small matrix size that comes with it does not allow
for many outlying eigenvalues.
At least our method 1 seems to be safe, being merely a time
average over correlations of the same set of stocks.
Another potential drawback is that the chopping procedure is
clearly not suitable for abrupt changes in the market, which would be diluted
in averaging over time windows.
Overall it is very interesting
to see how well a simple one-parameter power-law RMT can describe deviations
from the standard null-hypothesis of no correlations,
both for the global density and global spacing distribution.

\indent

\noindent
\underline{Acknowledgements}:
Financial support by European Community Network ENRAGE
MRTN-CT-2004-005616 (G.A.)
is gratefully acknowledged. We would like to thank Richard Hawkes
for sharing his data with us.


\begin{appendix}

\sect{Global Densities of Ensembles of Covariance Matrices vs RMT
}\label{macrochop}

In this section we display the global density of three different ensembles of
covariance matrices generated with method 1 and 2. The purpose is
twofold. First we confirm that the generalised model
eq. (\ref{Pgen}) gives a better fit to the global density than the standard
WL eq. (\ref{PWL}). In this way we can determine the power $\alpha$ for the
decay, that could be used in a comparison to local statistics.

\begin{center}
\begin{figure}[htb]
\centerline{
\epsfig{file=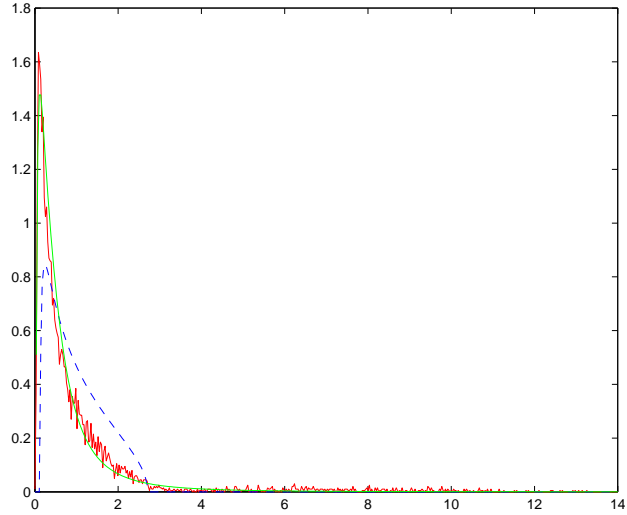,clip=,width=10cm}
}
\caption{Comparison between the full spectral density from an ensemble
generated using method 1 on set DJ ($48\times 20$ chopped ensemble).
The generalised density eq. (\ref{rhogen}) fitted with $\alpha=1$,
and the MP
distribution eq. (\ref{MP}) are given by the full green line and dashed blue
line, respectively.
}
\label{DJ2048rho}
\end{figure}
\end{center}

Second, we can
see a qualitative difference from the global density of a single covariance
matrix, fig. \ref{Comparison} in section \ref{data}, and fig.  \ref{DJ2048rho}
here: there are far fewer outliers here than there. In fact all points outside
the (crude) MP fit in  fig.  \ref{DJ2048rho} are due to the single largest
eigenvalue of various ensemble members, see fig. \ref{rawsmallR} right.
What could be reason for this? Of course the matrix sizes differ considerably,
$N=401$ in the unchopped set SP vs $N=20$ eigenvalues
in set DJ using method 1. We cannot exclude the
possibility that the chopping method reduces or washes out relevant
information about correlations.
In fact using method 2 we could somewhat expect that averaging over
submatrices from different stocks will reduce the information content.
\begin{center}
\begin{figure}[htb]
\centerline{
\epsfig{file=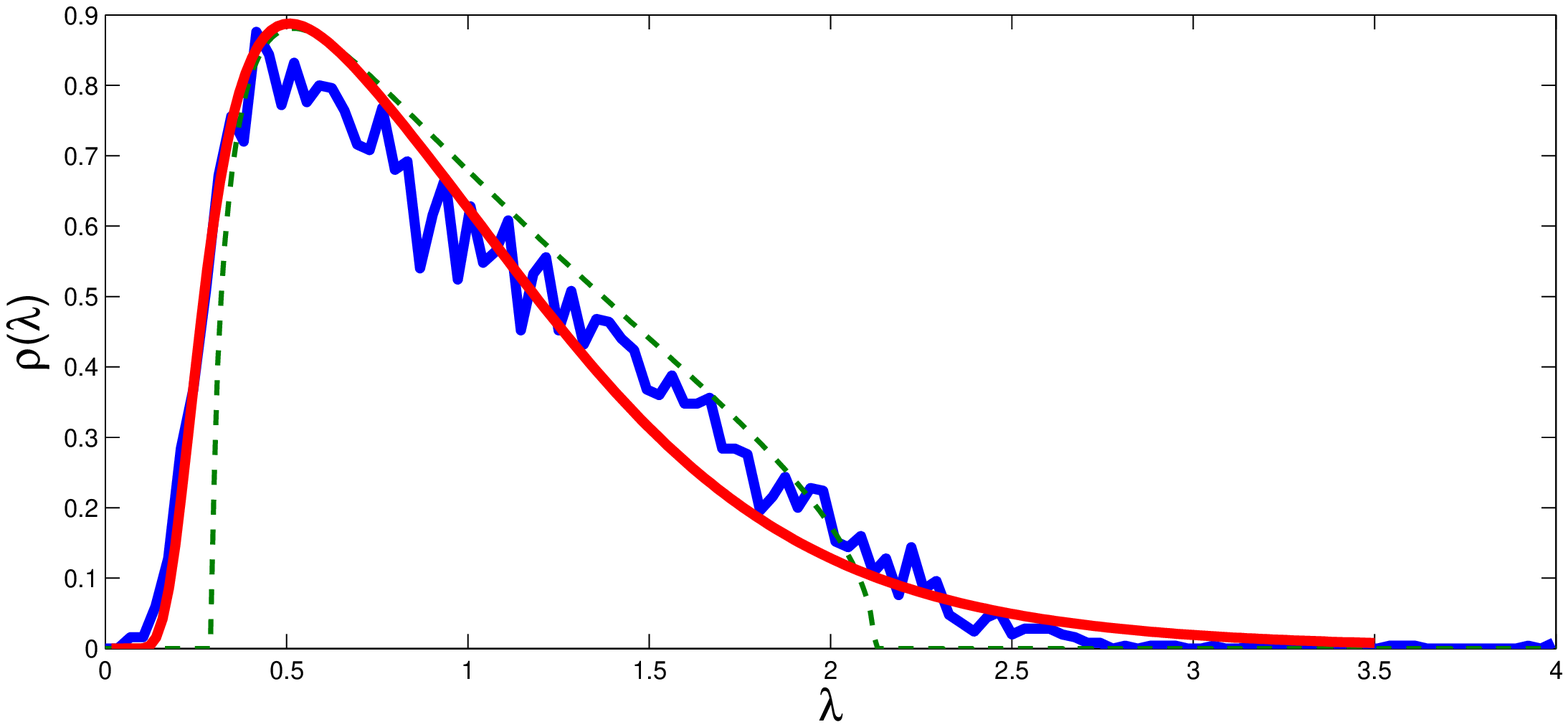,clip=,width=9.3cm}
\epsfig{file=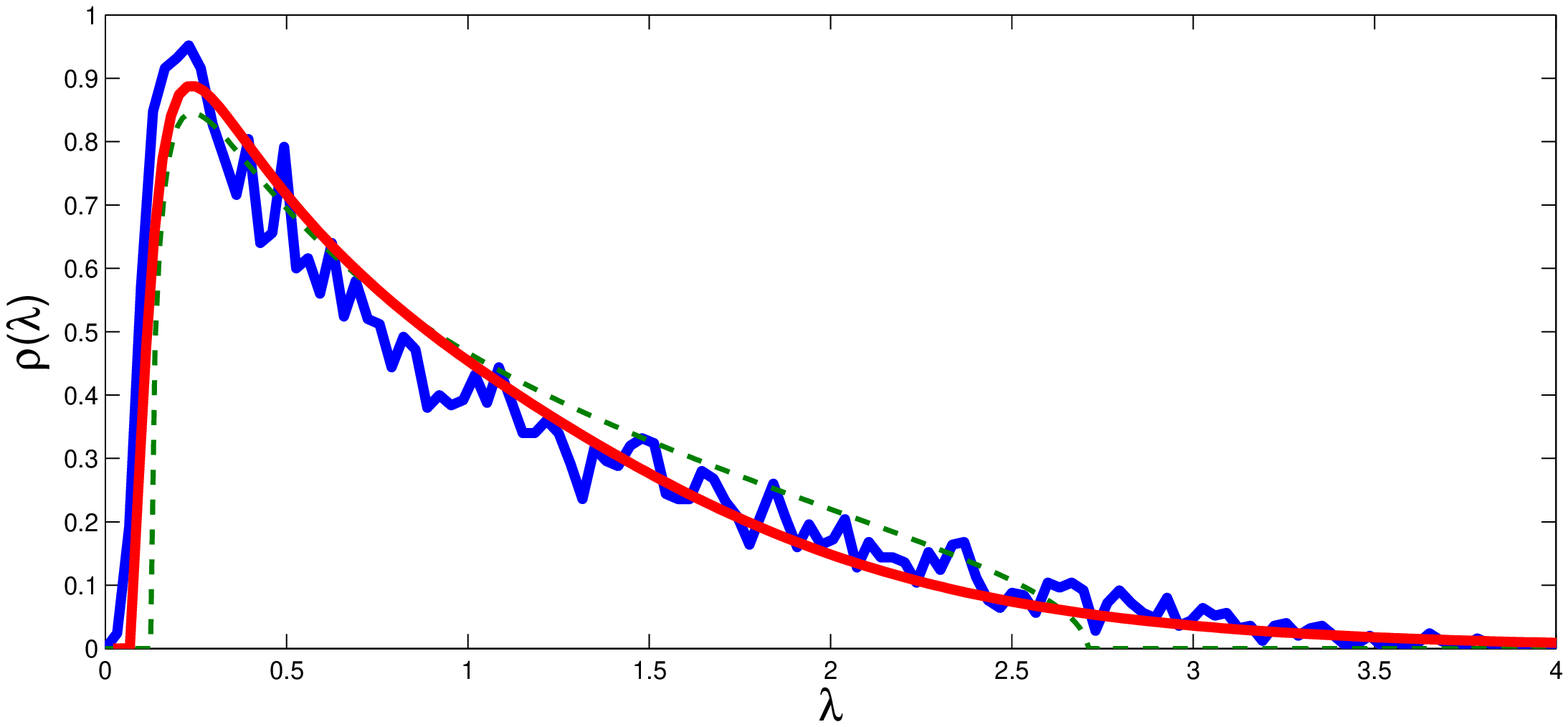,clip=,width=9.3cm}}
\caption{Comparison between part of the global spectral density from ensembles
using method 2 on data set SP:
the $48\times 10$ chopped ensemble (left) and $48\times 20$ (right) vs
the generalised density eq. (\ref{rhogen}) (full red line)
for $\alpha=7,8$ respectively. The MP
distribution eq. (\ref{MP}) (dashed line) gives a worse fit.}
\label{SPrho}
\end{figure}
\end{center}

\sect{Consistency of Chopping in Method 2: Permutations}\label{chopperm}

In this appendix we check the effect of taking different choppings in method 2
by comparing different ensembles when randomly
permuting the rows (stocks) into different
groups. We have created 10 different ensembles of the same size $48\times20$
from data set SP.
For each ensemble the 400 stocks were put into a different partition of 20
blocks of 
size 20.

\begin{center}
\begin{figure}[htb]
\centerline{
\epsfig{file=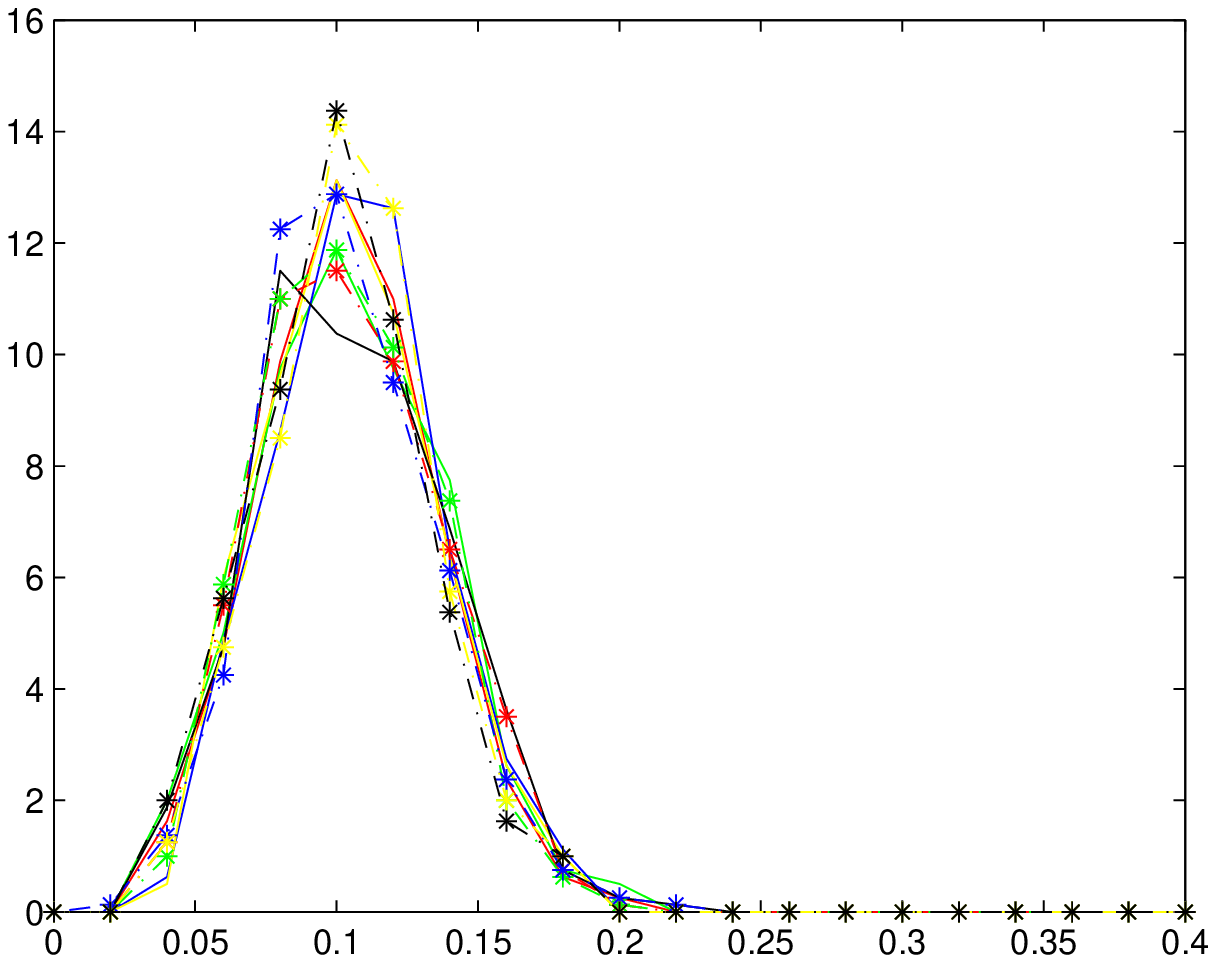,clip=,width=5.8cm}
\epsfig{file=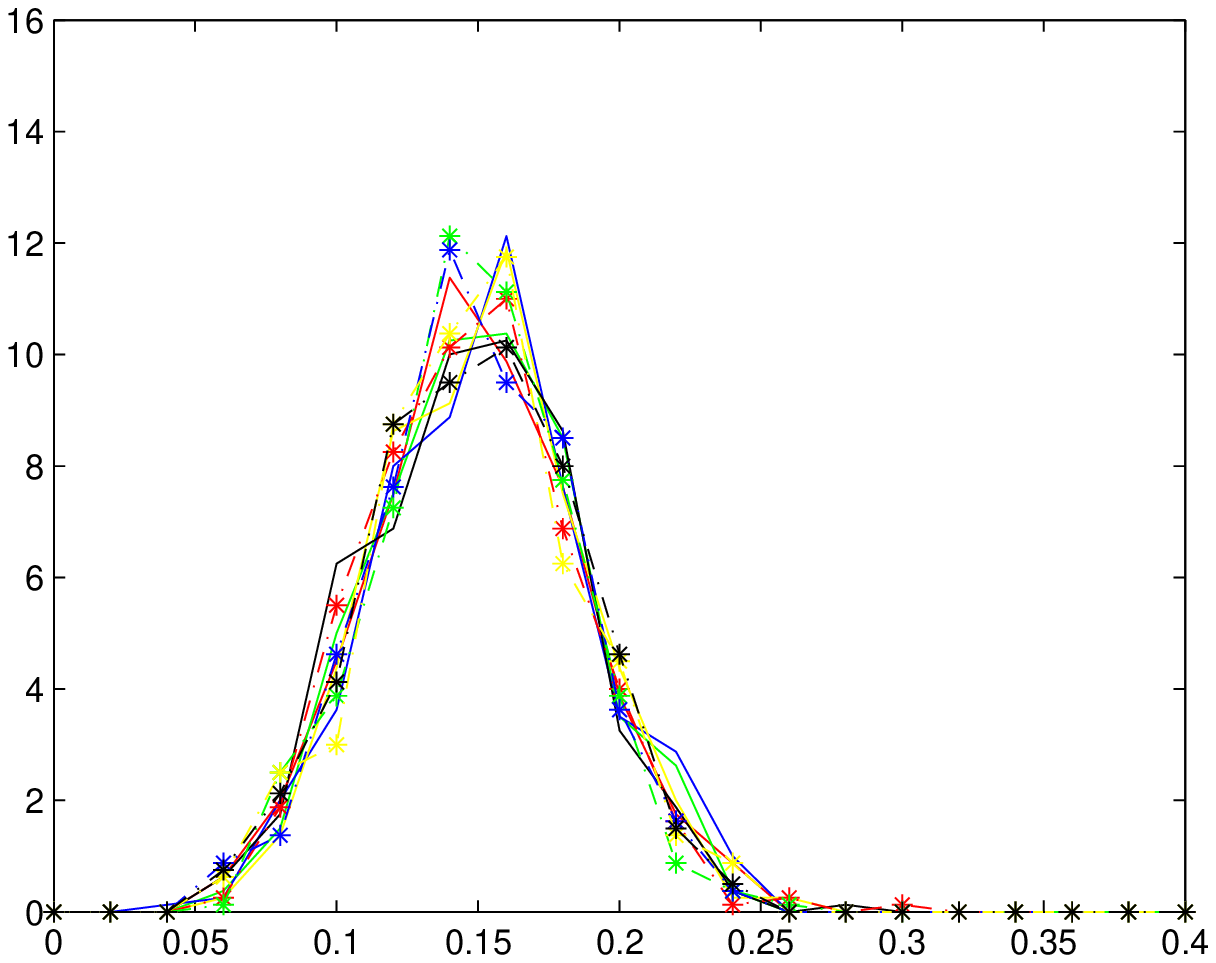,clip=,width=5.8cm}
\epsfig{file=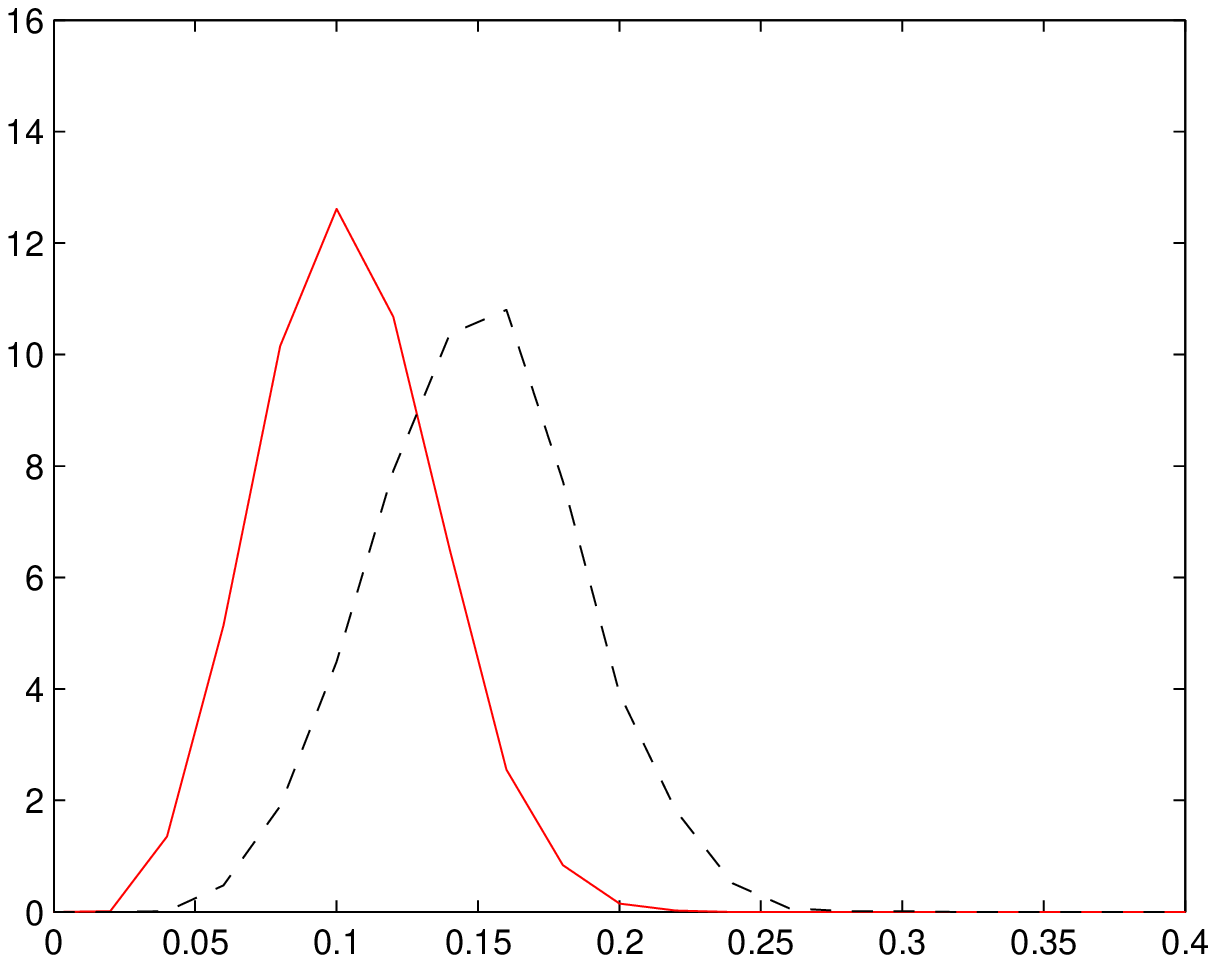,clip=,width=5.8cm}
}
\caption{
Consistency of the 2 smallest eigenvalues in method 2:
10 different choppings of the same size
$48\times20$ are obtained by using different partitions of the 400 stocks into
20 blocks of size 20:
the overlapping of the smallest (left) and second smallest eigenvalue (middle)
for these 10 different ensembles.
Right plot: after taking the average over the 10 ensembles the averaged
smallest (red left curve) and averaged second smallest eigenvalue (right
dashed curve)
are superimposed.
} \label{permsm}
\end{figure}
\end{center}

The superposition of the smallest  and second smallest eigenvalues
from these 10 different ensembles generated by method 2
is shown in fig. \ref{permsm}.
We find that both eigenvalues superimpose well to give a smoothed curve. For
illustration we display the curves of the two eigenvalues averaged over these
10 ensembles in fig. \ref{permsm} right, to illustrate
their width and separation
(compared to a single ensemble in fig. \ref{rawsmallR}).

\begin{center}
\begin{figure}[htb!]
\centerline{
\epsfig{file=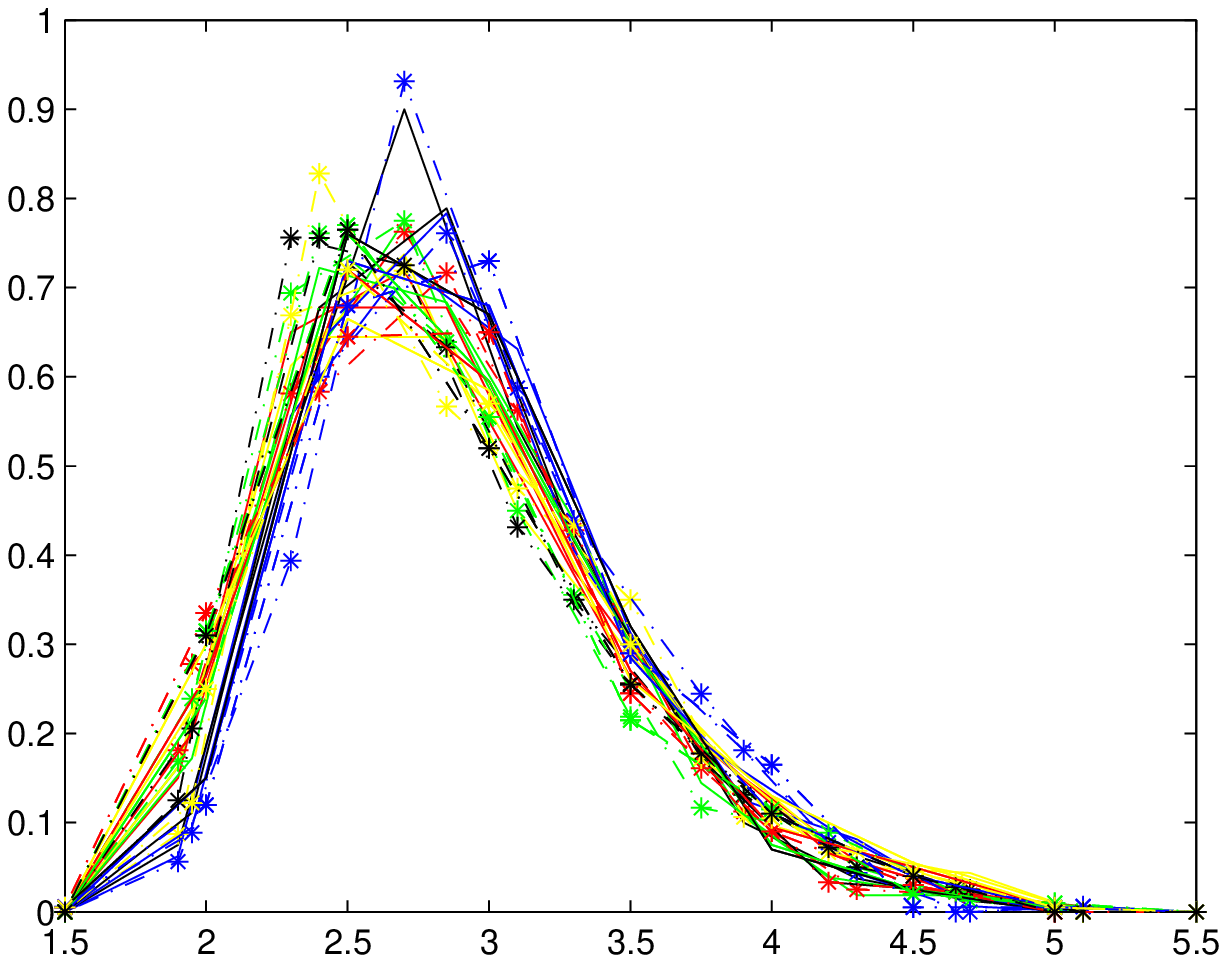,clip=,width=5.8cm}
\epsfig{file=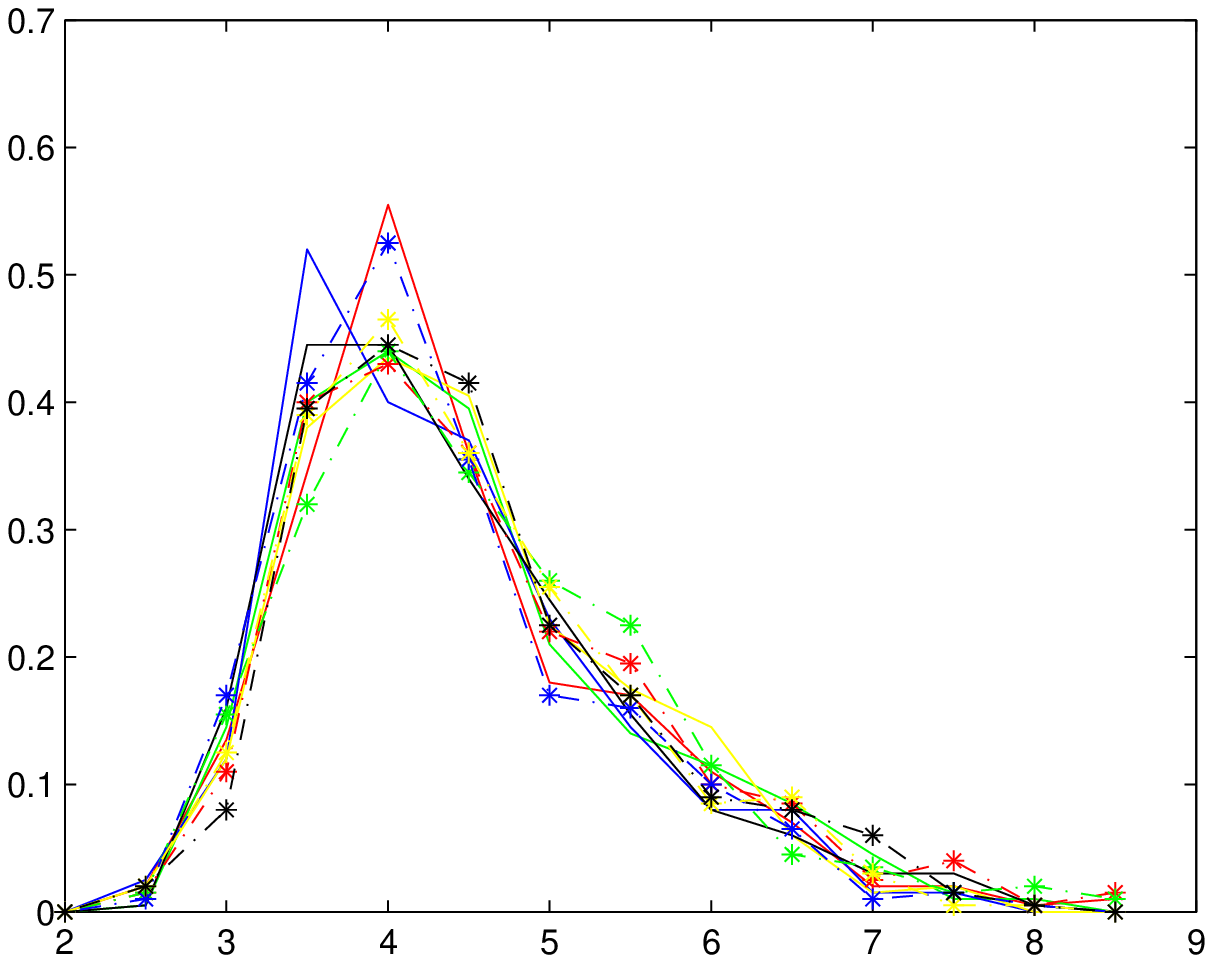,clip=,width=5.8cm}
\epsfig{file=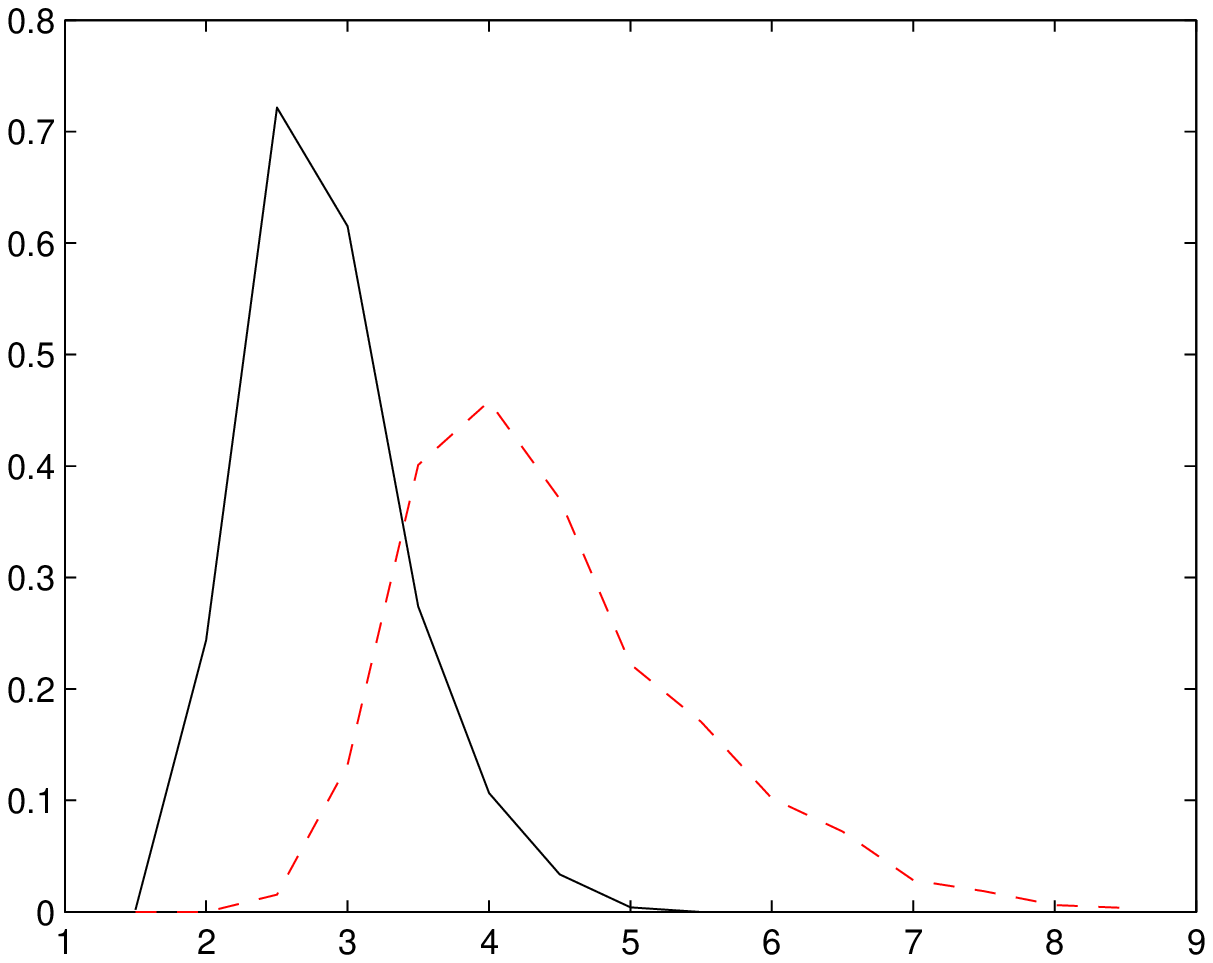,clip=,width=5.8cm}
}
\caption{
Consistency of the 2 largest eigenvalues in method 2 using the same method as
in fig. \ref{permsm}:
the overlapping the second largest (left) and largest eigenvalue (middle)
from 10 different ensembles.
Right plot: after taking the average over the 10 ensembles the averaged
second largest (black left curve) and averaged largest eigenvalue
(right dashed curve) are superimposed.
} \label{permlg}
\end{figure}
\end{center}

For curiosity we have done the same analysis for the largest and second largest
eigenvalues within the same 10 ensembles as shown in fig. \ref{permlg}.

As we can see in fig.  \ref{permlg}
the largest eigenvalues from the 10 different ensembles overlap smoothly.
This fact makes its unlikely that in these ensembles the largest eigenvalues
still represents the market movement as in
\cite{plerouplus,Plerouetal}. It would be very interesting to try to identify
the largest eigenvalues in fig.  \ref{permlg} as a largest
generalised TW eigenvalue from RMT as shown
in fig \ref{TWgen} right. However, due
to the lack of an analytic formula within our generalised RMT,
and in particular because of the (possibly
$\alpha$-dependent) scaling with $N$ this is a difficult task.

\sect{Unfolding Routine}\label{unfold}

Consider the cumulative distribution of eigenvalues:
\begin{equation}
P(x)=N\int_0^x d x^\prime\rho(x^\prime)
\end{equation}
where $\rho(\lambda)$ is the smooth density of eigenvalues (normalised to $1$).

Given a set of $m$ samples of $N$ bare eigenvalues
$\{\lambda_1^{(k)},\ldots,\lambda_N^{(k)}\}$ (with $k=1,\ldots,m$),
we first define a set of unfolded eigenvalues $\{E_1^{(k)},\ldots,E_N^{(k)}\}$
as:
\begin{equation}
E_j^{(k)}=P(\lambda_j^{(k)})
\end{equation}
Then the set of spacings $\{s\}$ to be histogrammed is computed by the
nearest-neighbour difference among the $E$'s \emph{within}
each sample. Clearly, for empirical data the main problem is to estimate
reliably the cumulative distribution $P(x)$.
Given a regular grid of $K$ points $\{0\leq y_1\leq\ldots\leq y_K\}$ on the
positive semi-axis, one can simply
define an estimator $\tilde{P}(y_r)$ for $P(x)$ as  the number of bare
eigenvalues (in total $mN$)
falling below $y_r$, divided by $m$. This estimator approaches $N$ for large
argument $y$. Then, in order to get a
continuous distribution, one simply performs a polynomial fit of degree $d$ on
the $\{y_r,\tilde{P}(y_r)\}$ pairs and approximates the
true cumulative distribution with the fitting polynomial $p_d(x)\approx P(x)$.

The \textrm{MATLAB} code that performs the unfolding procedure as stated
above and returns the positions $X$ of the bins
and the normalised histogram $Y$ of the nearest neighbour spacings can be
retrieved from the public domain web site \cite{pier}.

\end{appendix}

\end{document}